\documentclass[11pt]{article}

\usepackage{epsfig,latexsym}

\newcommand{\sect}[1]{\setcounter{equation}{0}\section{#1}}
\renewcommand{\theequation}{\arabic{section}.\arabic{equation}}

\def\al {\alpha}

\def\ba{\begin{eqnarray}}
\def\bb{{\cal B}}

\def\be{\begin{equation}}
\def\bt {\beta}
\def\br{{\bf r}}
\def\Bar {\overline}
\def\cc{{\cal C}}

\def\de{\delta}
\def\dd{{\cal D}}
\def\di{\partial}

\def\ea{\end{eqnarray}}
\def\ee{\end{equation}}

\def\half{\frac{1}{2}}

\def\la {\lambda}
\def\lll{\left(}

\def\LLL{\left[}

\def\nnn{\noindent}
\def\om{\omega}
\def\oo{\over}
\def\Om {\Omega}

\def\pp{\cal P }

\def\rr{{\cal R }}
\def\rrr{\right)}

\def\RRR {\right]}
\def\si{\sigma}
\def\td{\tilde}

\title{{\bf Thermodynamics and Self-Gravitating Systems}}
\author{ Joseph Katz
 \\ {\it The Racah Institute of Physics, Givat Ram, 91904 Jerusalem,
Israel.}\\{\it email: jkatz@phys.huji.ac.il}}

\date{ }

\begin{document}
\maketitle
\centerline{{\bf ABSTRACT}}

 This work assembles some basic theoretical elements
on thermal equilibrium,  stability conditions,  and fluctuation
theory  in self-gravitating systems illustrated with a few examples.
Thermodynamics deals with  states that have settled down after
sufficient time has gone by.  Time dependent phenomena are beyond the
scope of  this  paper. While thermodynamics is firmly rooted in
statistical physics, equilibrium configurations, stability criteria
and the destabilizing effect of fluctuations  are all expressed in
terms of thermodynamic functions. The work  is not a review paper but 
a pedagogical introduction which may interest theoreticians in
astronomy and astrophysicists.  It contains  sufficient mathematical
details for the reader to  redo all  calculations. References are only
to  seminal works or readable reviews\footnote{Except in the ultimate
sub-section.}. Delicate mathematical problems are mentioned but are
not discussed in detail.
Statistical mechanics with long range forces.
\vskip 0.5 cm
\centerline{{\it Dedicated to my late friend Professor Gerald Horwitz
(1929-2001)}}


\sect{Introduction}

 A strident aspect that puts  self-gravitating systems in a class of
their own   is that they cannot be small sub-systems  of a large
ensemble except in very imaginary situations.   A small sub-system of a
large self-gravitating system  has its energy almost entirely 
determined  by the large system. Moreover, the sum of  energies of all
the small sub-systems is  not equal to  the total energy of the system
in contrast with a basic tenet of classical thermodynamics. This has
far reaching consequences. Much of the rich harvest of classical and
quantum thermodynamics as exposed in Landau and Lifshitz's {\it
Statistical Physics} \cite{LL}  is useless in gravitational
thermodynamics.  For instance stable and unstable isolated
self-gravitating systems   may have  heat capacities of both
signs. This is also the case for a system 
 in a heat bath. Thus
$C_V>0$ is no more a signature of  stable thermal states like in
classical  thermodynamic ensembles. In this respect gravitational
thermodynamics appears slightly exotic though not to astronomers as
pointed out  by  Lynden-Bell \cite{LB} in his  review on progress in
understanding of gravitational thermodynamics. It is also what makes  
thermodynamics of self-gravitating systems  interesting.

The  paper is based on the Gibbs density of states function in the
meanfield approximation.  Astronomers and astrophysicists are more
familiar with Boltzmann entropy. Gibbs definition is more general in
that it provides Boltzmann entropy as a first approximation as well as
stability conditions and fluctuations. It has a wider range of
applications and extends to meanfield theory. The Gibbs approach is in
our view the least frustrating way to understand  statistical
thermodynamics and  provides the best logically connected  approach to
the subject.  

Section 1 begins   with a review of  statistical equilibrium theory in
the simple case  of $N$ identical point masses  in an external field
first and gravitationally interacting particles in a finite volume
next.   The case
$N=2$ 
 is  interesting and has been studied in detail by Padmanabhan
\cite{Pad}.   Here we consider $N>>2$  and situations in which 
particle pair formation has a negligible role on relevant time scales. 
In this case   meanfield theory is a very good approximation.  A
steepest descent calculation gives equilibrium configurations of 
stable as well as metastable states.  Metastable states are
particularly  interesting  in gravitational thermodynamics\footnote
{Most globular clusters
\cite{MH} and rich clusters of galaxies
\cite{Sa} have their cores in  a quasi thermalized metastable state.}.
The method  provides also stability conditions.

In section 2 we deal with thermodynamic functions  and stability
conditions in general. The latter are  first examined with the help of
Gibbs' density of states  function  but are subsequently translated
into conditions involving only  thermodynamic functions. The theory 
applies  to
 isolated non-extensive  systems, the type of system considered in
section 1, but
 is of  much  broader applicability. It predicts in particular
that under very general assumptions   {\it stable} thermal equilibrium
configurations of isolated systems {\it near} instability have always
negative heat capacities which turn to be positive  when   systems
become unstable! In classical thermodynamics   a small thermally
stable subsystem of an ensemble has always a positive heat capacity. 
 These results imply  that stable isolated systems near instability
are necessarily  unstable if put in a heat bath. This
well known ``nonequivalence of ensembles" has  no analogue in
classical  thermodynamics. 

Section 3 develops the theory of fluctuations. We deal with   
fluctuations of  temperature as a generic example. Fluctuations of
other thermodynamic functions can   be dealt with in a similar way.
The theory has also broad applicability;  again results are expressed
in terms of thermodynamic functions  and hold only  near instability.
For nonequivalent ensembles the behavior of equilibrium  near
instability  departs considerably from classical systems  and the
effect of fluctuations is consequently quite different. 

 Section 4 gives  two examples of application with some detail: 
isothermal spheres in various ensembles and liquid ellipsoids. The
purpose here is to illustrate the power and limitations of
thermodynamics
 in self-gravitating systems. The last sub-section reviews  briefly
various other applications in astronomy where thermodynamics has been
used or might have been with considerable benefit. 

Each section has its own summary, conclusion and/or comments.


\sect{Statistical thermodynamics}

It may   interest
 the reader to see the explicit  connection between
the entropies of Gibbs and  Boltzmann  by starting with the simpler
case of a system in an  external field first. Here some of the
mathematical techniques appear in a more transparent form. We go then
over to  self-gravitating systems in the meanfield approximation which
is the main object of this section.
\vskip .5 cm


\noindent\normalsize{\it (i) Gibbs's density of states function and 
Boltzmann entropy.}

\vskip .5 cm

Consider first the simple case of $N$ identical point particles  of
mass $m$  in a 
 volume $V$, with  perfectly reflecting massless walls in an external
gravitational potential\footnote {To avoid any confusion  we use here a tilde over all   quantities
with the same symbols and the  same meanings as in  self-gravitating
systems.}
$\td U({\bf r})$. The total energy  $\td H$, the sum of kinetic
$\td H_K$ and potential energy
$\td H_P$ of the system is a constant of motion. Let $\td E$ be its
value.  The total energy  $\td H$ is a function of the coordinates  of
each particle in phase space  $({\bf r}_i,{\bf p}_i )$,
$(i=1,2,...,N)$:
\be
\td H= \td H_K+\td H_P=\sum\limits_{i} {1\over
2m}p_i^2+\sum\limits_{i} m\tilde U({\bf r}_i)=\sum\limits_{i}\td
E_i=\td E.
\ee  The total energy  $\td H$ is here equal to the sum of the energies
of individual particles $\td E_i$. The density of states function
$\td\Om (\td E, V, N, m)$ or in short
$\td\Om (\td E)$ is the  sum of all possible states\footnote  {The
Gibbs density fo states function  has dimensions $[\td\Om]
=erg^{3N-1}sec^{3N}$.  Thus
$\td\Om$ like later $\Om$ is defined up to a constant that depends on
the units chosen.} of the system with energy
$\td E$ divided by
$N!$:
\be
\td\Om (\td E)={1\over N!}\int \delta(\td E-\td H)\prod\limits_i
d\om_i~~~{\rm where} ~~~d\om_i=d^3r_id^3p_i~~~(-\infty <{\bf
p}_i<+\infty);
\label{Om11}
\ee
$\delta(\td E-\td H)\prod\limits_i d\om_i$  is known as the
microcanonical distribution
\cite{LL} and the basic justification for this distribution is  the
ergodic hypothesis. The Gibbs entropy
$\td S_g$   is the logarithm of $\td\Om(\td E)$ divided by 
Boltzmann's  constant $k$. Here we take
$k=1$ which amounts mainly to measure the temperature in units of
energy.
 Thus
 \be
 \td\Om(\td E)=e^{\td S_g}.
 \ee 
 To evaluate the Gibbs entropy we  first replace 
$\delta(\td E-\td H)$ by a Bromwich  integral of $1$ (its inverse
Laplace transform) in brackets:  
\be
\td \Om (\td E)={1\over N!}\int\LLL {1\over 2\pi
i}\int_{b-i\infty}^{b+i\infty} e^{\beta(\td E-\sum\limits_{i}\td
E_i)}d\beta\RRR\prod\limits_i d\omega_i.
\label{Om12}
\ee

This expression for $\td\Om$  can be rewritten  after a slight 
rearrangement of terms as
\be
\td\Om(\td E)={1\over 2\pi i }\int_{b-i\infty}^{b+i\infty} \LLL{1\over
N!}\int
\prod\limits_i e^{-\beta \td E_i}d\omega_i \RRR e^{\beta \td E}d\beta
\equiv {1\over 2\pi i }\int_{b-i\infty}^{b+i\infty} \td\Psi e^{\bt \td
E} d\beta ,
\label{Om13}
\ee
$\td\Psi$, the quantity between brackets,   is easily seen to reduce
to
\be
\td\Psi={1\over N!}\left(\int e^{-\beta  \td E_0}d\om\right )^N~~~{\rm
where}~~~ \td E_0={1\over 2m}p^2+m\tilde U({\bf r})~~~{\rm and}~~~
d\om=d^3rd^3p.
\label{Psi11}
\ee
$\td E_0$ is the energy of of one particle with coordinates $({\bf r},
{\bf p})$. We can of course also write $\td\Psi$ as follows
\be
\td \Psi=\sum\limits_{N'=0}^{N'=\infty}{1\over N'!}\left(\int
e^{-\beta 
\td E_0}d\om\right )^{N'}\delta_{N'N}.
\ee  We then approximate $\delta_{N'N}$ by $\delta(N'-N)$ since $N>>1$
and replace  immediately
$\delta(N'-N)$ by a Bromwich integral like we did with $\delta(\td
E-\td H)$:
\be
\td\Psi\simeq {1\over 2\pi
i}\int_{a-i\infty}^{a+i\infty}\sum\limits_{N'=0}^{N'=
\infty}{1\over N'!}\left(
\int e^{-\beta \td E_0}d\om\right )^{N'} e^{\al(N'-N)}d\al.
\label{tildePsi1}
\ee
$\td\Psi$ can now be rewritten in this form
\be
\td \Psi\simeq{1\over 2\pi i}\int_{a-i\infty}^ {a+i\infty}e^{\lll-\al
N+\int \tilde fd\om\rrr}d\al~~~\Rightarrow~~~\tilde f=e^{\al-\beta 
\td  E_0}. 
\label{Psi12}
\ee and if we  substitute this expression for $\td\Psi$ back into
(\ref{Om13})  we obtain 
\be
\td \Om(\td E)\simeq{1\over (2\pi
i)^2}\int_{a-i\infty}^{a+i\infty}\int_{b-i\infty}^ {b+i\infty}
e^{\tilde \sigma}d\al d\beta  ~~\Rightarrow~~\tilde \sigma=-\al
N+\beta \td E + \int \tilde fd\om.
\label{Om14}
\ee We shall now evaluate $\td \Om(\td E)$ by a steepest descent
technique \cite{Arf} as follows. We first look for the extremum of
$\td\si(\al,\bt)$; this is obtained from the following equalities:
\be  {\di\td\si\over \di\al}=-N+\int \tilde fd\om=0~~~,~~~
{\di\tilde\sigma\over
\di\bt}=\td E-\int \td E_0\tilde fd\om=0.
\label{NE11}
\ee The equations define a point $(\td\al_e,\td\bt_e)$ in $(\al,\bt)$
space in terms of
$\td E$ and
$N$ and a corresponding equilibrium value $\tilde f_e$ for $\td f$:
\be
\tilde f_e=e^{\td\al_e-\td\beta_e \td E_0}=e^{\td\al_e-\td\beta_e
[{1\over 2m}p^2+m\tilde U({\bf r})]}.
\label{tildef}
\ee
 $\td f_e$ is the Boltzmann distribution of energies in the external
field $\td U({\bf r})$ calculated at the point of extremum. The
 extremal value
$\td\si_e$ of
$\td\si$ is the Boltzmann entropy 
$\td S$ and it is easily seen that
\be
\td S=\tilde \si_e=-\int \tilde f_e\ln(\tilde f_e)d\om+N.
\label{tildeS}
\ee
 We may thus write (\ref{Om14}) as follows introducing 
$\delta^2\td \si$ which represents the sum of all the terms of order
higher than one in a Taylor expansion of $\tilde\si$ near the extremum 
$\td\si_e$:
\be
\tilde\Om(\td E)=e^{\tilde S_g}\simeq e^{\tilde S}{1\over (2\pi
i)^2}\int_{a-i\infty}^{a+i\infty}\int_{b-i\infty}^{b+i\infty}
e^{\delta^2\tilde \si}d\al d\beta . 
\label{Om15}
\ee 
Since the exponent $\delta^2\td \si$ is of order $N$ 
the integrant in (\ref{Om15}) is very steep and a good approximation
is  obtained by  limiting the integrations to terms of order two. The
terms of order two are readily found; with $\delta
\al=\al-\td\al_e$ and
 $\delta\bt=\bt-\td\bt_e$ we may write
\ba 
{\delta^2\td \si}&\simeq& \half \LLL
\lll{\di^2\td\si\oo\di\al^2}\rrr_e(\delta
\al)^2+2\lll{\di^2\td\si\oo\di\al\di\bt}\rrr_e\delta
\al\delta \bt+\lll{\di^2\td\si\oo\di\bt^2}\rrr_e(\delta
\bt)^2\RRR,\nonumber\\ &=&\half \LLL N(\delta \al)^2
-2E\de\al\de \bt+N\Bar{\td E_0^2}(\delta
\bt)^2\RRR\nonumber\\ &=&\half N\LLL  (\de\al-\Bar{\td
E_0}\de\bt)^2+\Bar{(\td E_0-\Bar{\td E_0})^2}\de\bt^2\RRR ,
\label{delta2si}
\ea 
in this formula mean values calculated in the phase space of a
particle are denote with an overbar; for instance:
\be
\Bar{\td E_0}={1\oo N}\int \td E_0 \td f_e d\om
\ee We  can now fix the position of the imaginary axis for the complex
variables
$(\al,\bt)$ by introducing  a new set of variables which vary between
$\pm
\infty$:
\be
\td \al^*=i(\delta \al-\Bar{\td E_0}\delta \bt) ~~~{\rm and}~~~\td
\bt^*=i\delta \bt.
\label{idelta}
\ee 
This gives to (\ref{Om14})  the following form:
\be
\td\Om(\td E)=e^{\td S_g}\simeq e^{\td S}{1\over (2\pi
)^2}\int_{-\infty}^{+\infty}e^{-{N\oo
2}\tilde \al^{*2}}d\tilde \al^*\int_{-\infty}^{+\infty}e^{-{N\oo
2}\Bar{(\td E_0-\Bar{\td E_0})^2}\tilde\bt^{*2}}d\tilde\bt^*.
\label{tdOm}
\ee 
The integrals are easily found and the end product is this:
\be
\td S_g\simeq \td S -\ln(N) -\ln(2\pi)- \half\ln\LLL~\Bar{(\td
E_0-\Bar{\td E_0})^2}~\RRR.
\ee This shows incidentally that $\td S$ is a maximum of $\td\si$ for
any $\td U$; in other words the system is stable and $\td\Om(\td E)$
is convergent.  Moreover, since $\td S$ is of order $N$ and
$N>>1$ the last three terms are clearly negligible compared to $\td S$
so that
\be
\tilde S_g\simeq\tilde S.
\ee
This is the main result of this first subsection.


\vskip .5 cm
\noindent\normalsize{\it (ii) The Gibbs density of states function for
self-gravitating systems.}

\vskip .5 cm

If the same system of $N$ particles of mass $m$ in a volume $V$ are
interacting gravitationally, the total conserved  energy is
\be H=H_K+H_P=\sum\limits_i {1\over 2m} p_i^2-{1\over 2}
\sum\limits_{i\neq j}{Gm^2\over | {\bf r}_i-{\bf r}_j|} = E.
\label{H1}
\ee
 $G$ is the gravitational constant. The density of states function
$\Om (E)$ is the same as (\ref{Om11}) without  tildes:
\be
\Om (E)={1\over N!}\int \delta(E-H)\prod\limits_i d\om_i.
\label{Om1}
\ee As is well known, $\Om$ diverges.  This can easily  be seen as
follows\footnote{Here we follow Padmanabhan \cite{Pad}.}. Take
equation (\ref{Om1}), replace again $\delta(E-H)$ by a Bromwich 
integral  of
$1$ with variable $E-H=E-(H_K+H_P)$; in this way we obtain an
expression similar to (\ref{Om12}) :
\be
\Om (E)={1\over N!}\int\LLL {1\over 2\pi i}\int_{b-i\infty}^{b+i\infty}
e^{\beta(E-H_K-H_P)}d\beta\RRR\prod\limits_i d\omega_i.
\label{Om3}
\ee 
Integrate $e^{-\bt H_K}$ over  momentum space which is easy to do; 
this gives 
\be
\Om (E)={(2\pi m)^{{3\over 2}N}\over N! }\int_V\LLL {1\over 2\pi
i}\int_{b-i\infty}^{b+i\infty} {e^{\beta(E-H_P)}\over
\bt^{{3\over 2}N}}d\beta\RRR\prod\limits_i d^3r_i.
\ee Then use Table 15.2 in Arfkin \cite{Arf} to calculate the Bromwich
integral in the square brackets and replace $H_P$ by its definition in
(\ref{H1}):
\be
\Om (E)={(2\pi m)^{{3\over 2}N}\over N! ({{3\over 2}N}-1)!}\int_V
\left(E+{1\over 2}\sum\limits_{i\neq j}{Gm^2\over | {\bf r}_i-{\bf
r}_j|} \right)^{{3N\over 2}-1}
\prod\limits_id^3 r_i,
\ee and finally change to new variables ${\bf r}={\bf r}_1-{\bf r}_2$
and ${\bf r}_u$ with
$u,v=2,3,...,N$ in terms of which
\be
\Om(E)={(2\pi m)^{{3\over 2}N}\over N! ({{3\over 2}N}-1)!}\int_V {\cal
I }\prod\limits_u d^3r_u,
\ee in this equality,
\be {\cal I}=\int_V \left( E+{Gm^2\over
r}+\sum\limits_{u=3}^{u=N}{Gm^2\over |{\bf r}+{\bf r}_2-{\bf
r}_u|}+{1\over 2}\sum\limits_{u\neq v}{Gm^2\over |{\bf r}_u-{\bf
r}_v|}\right)^{{3N\over 2}-1}d^3r.
\ee We  now  see  that for  $r\rightarrow 0$, ${\cal
I}\rightarrow\propto\int r^{3-{3N\over 2}}dr\propto r^{4-{3N\over
2}}$. So for
$N\geq3$ and  $r\rightarrow 0$,  
$\Omega\rightarrow\infty $. Considerations of simple models show that
the system  with a
 subset of particles closely bound together by gravity and the
remainder banging around with the high energy released has a large
phase space volume associated with it \cite{LBW}. Such a state can
presumably be
 reached given sufficient time for the system to evolve. A system of
point particles is however not realistic.  In more realistic models,
one make  use of small hard spheres or fermions or remove the
divergence of the  potential   with artificial cutoffs. This makes
$\Om(E)$ converge. 

It is important to realize that once $\Om$
converges by say taking hard spheres, the stable state is
not necessarily  a dense core with a dilute halo; other configurations
may be more stable. For a sufficiently small cut off at least two
particles will be very close together since such states would dominate
were the hard spheres reduced to points. Other configurations may only
be more stable if the cut offs are big enough.

Models like these have been  reanalyzed recently
by Chavanis
\cite{Chav} who reviews earlier interesting works with various 
cutoffs. Neither of the models give however a direct evaluation of
$\Om(E)$.


\vskip .5 cm
\noindent\normalsize{\it (iii) The meanfield approximation.}
\vskip .5 cm

A meanfield theory is a  most common 
approximation when short distance effects are negligible on the
timescales considered. A steepest descent calculation gives a
meaningful statistical thermodynamics, equilibrium configurations and
stability conditions.

The following calculation is kept simple by not introducing any sort
of cut off while embedding divergences into  terms that would
normally converge if there was a cutoff. These quantities when finite
have no role in the  results that interest us.

One should however be aware that the maximum of the new integrand
$e^\si$ is not unique nor is it necessarily the highest. A second
maximum  may be much higher if the hard spheres are small enough.
However, the calculation of that maximum is precisely brushed under the
rug.  A correct calculation is far too difficult. Our entropy will be
associated with what is often referred to as a ``local" maximum. This local
maximum may not be the highest but it may be the more important one
on  time scales relevant in astronomy.

The calculation of  $\Om(E)$ in  the meanfield approximation is
slightly intricate. We give here a heuristic  derivation  that  will
make the answer plausible. A   complete and rigorous derivation that 
follows paper \cite{HK} is given in Appendix A. 

Let us ask how  (\ref{Om14}) should be modified when $\tilde U({\bf
r})$ is replaced by the
gravitational field of the  particles themselves? The
field  is now  variable with $\infty^3$ degrees of liberty.
The correct
$\Om$ will have a functional integration over the space of all fields
instead of  a discrete integration over  the ${\bf r}_i$-space.
$\tilde \si$ needs a correction also that must depend on the field
because the mean value of the energy in equation (\ref{NE11}) is not
the correct expression for a a self-gravitating system: the potential
energy is counted twice. On the other hand the correction of $\tilde
\si$ should  not depend  on  $\al$ or $\bt$ since
$\td S$ must also be the correct expression for the Boltzmann entropy
whether the field is a self-field or an external one; the local
distribution does not know what the  origin of the field is.

With these remarks in mind, we now write what is the 
expression for the Gibbs density of states function $\Om$  in the
``meanfield approximation" derived from (\ref{Om3}) in Appendix A. First
we must use a field variable  $W$  slightly different from and more
convenient than
$\tilde U$. We replace $\tilde U$ by $ \beta^{-1/2}W({\bf r})$. In
terms of
$W({\bf r})$
\be
\displaystyle \Om(E)={1\over (2\pi
i)^2\bb}\int_{-\infty}^{+\infty}\dd
W\int_{a-i\infty}^{a+i\infty}\int_{b-i\infty}^{b+i\infty}
e^{\sigma}d\al d\beta~,
\label{Om4}
\ee in this expression $\bb$ is a constant which actually diverges if
there is no short distant cutoff (see appendix A).
${\cal D}W$ is the element of volume in function space (see also
appendix A) while
\be
\displaystyle \sigma[\al,\bt,W({\bf r})]=-\al N+\beta E +\int
e^{\al-\beta [{1\over 2m}p^2+m\beta^{-1/2}W(\bf {r})]}d\om+{1\over 8\pi
G}\int_{-\infty}^{+\infty} W\Delta W d^3r~.~~
\label{si}
\ee
$\Delta$ is the Laplacian. $\sigma$ differs from $\tilde \sigma$ in
(\ref{Om14})  mainly by the  term  
$W\Delta W$. This addition provides  the  correct mean value of the
total energy  as well as Poisson's equations for  equilibrium 
configurations as we shall now see. And it does not depend on $\al,
\bt$.


\vskip .5 cm
 {\it (iv)  A steepest descent evaluation}.
\vskip .5 cm
 We start as in the case with
an external field by calculating the extremal values of
$\si[\al, \bt, W({\bf r})]$; these are associated with equilibrium
configurations. 

The equilibrium value
$W_e({\bf r})$ of $W({\bf r})$ or rather   $U({\bf
r})=\beta_e^{-1/2}W_e({\bf r})$ is defined by 
\be {\delta \si\over \delta W({\bf r})}= 0~~\Rightarrow ~~\Delta U
=4\pi G\rho~~~{\rm with}~~~\rho({\bf r})=
m\int_{-\infty}^{+\infty}e^{\al_e-\bt_e [{1\over 2m}p^2 +mU({\bf
r})]}d^3p,
\label{Poisson}
\ee
$\rho$ is the mean mass density of the particles in a meanfield  $U$ 
with a Boltzmann distribution
\be f_e=e^{\al_e-\bt_e [{1\over 2m}p^2+mU({\bf r})]}=e^{\al_e-\bt_e
E_0},
\label{f}
\ee
$E_0$ is here the energy of one particle in the meanfield $U$ itself, 
a solution of Poisson's equation (\ref{Poisson}). The equilibrium
values $(\al_e,\bt_e)$ of $(\al,\bt)$  are  fixed by equations similar
to (\ref{NE11}) but $E$ is not quite the same as $\td E$:
\be {\di\si\over\di \al}=0\rightarrow N=\int f_e d\om   ~~~{\rm
but}~~~{\di\si\over\di \bt}=0\rightarrow E=\int
\left[{1\over 2m}p^2+{1\over 2}mU({\bf r})\right]f_e d\om.
\label{NE}
\ee The maximum $\si_e$ of $\si$, which is approximately the maximum of
the integral when $N>>1$ is equal to the Boltzmann entropy. It
is indeed easy to see that  (\ref{f})(\ref{NE}) provide the same 
expression for the extremal value of
$\si$ as (\ref{si}):
\be
 S(E,N,V)\simeq \si_e =-\int f_e\ln (f_e)d\om+N.
\label{S}
\ee

We may thus write (\ref{Om4}) as follows introducing  $\delta^2 \si$
which represents the sum of all the terms of order higher than one in a
Taylor expansion near the extremum:
\be
\Om(E)=e^{ S_g}\simeq e^{S}{1\over (2\pi
i)^2\bb}\int_{-\infty}^{+\infty}\dd
W\int_{a-i\infty}^{a+i\infty}\int_{b-i\infty}^{b+i\infty} e^{\delta^2
\si}d\al d\beta . 
\label{Om18}
\ee Since, as before, the exponent is of order $N$ the exponent in the
integral is very steep and a good approximation will be obtained by 
limiting the integration to terms of order two. The reader interested
in the derivation  of $\delta^2 \si$  and its explicit expression will
find  them both  in Appendix B. There we show that $\Om(E)$ is
approximately given by
\be
\Om(E)\simeq\ e^{S-\ln(N) -\ln( 2\pi b\bb/\bt_e)}  
\int_{-\infty}^{+\infty}e^{\delta^2\si'}\dd (\delta W),
\label{Om19}
\ee
$\delta^2\si'$ is a non-local quadratic functional in $\delta W$ 
which must be negative for the integral to converge, something of the
form:
\be
\delta^2\si'=\int_{-\infty}^{+\infty}\delta W{\cal O}(\br,\br ')
\delta W' d^3r d^3r'<0.
\label{O}
\ee
 $\delta W$ can in principle  be expanded in terms of a complete set of
orthonormal eigenfunctions say $\xi_a(\br)$ of the linear operator
${\cal O}(\br,\br ')$ with arbitrary discrete variables
$\delta Y^a$. The quadratic form $\delta^2\si'$ is then replaced by an
infinite sum of squares
$(\delta Y^a)^2$ with coefficients that for convenience we write
$-\half\la_a'$; thus
\be
\delta^2\si' = -\half\sum\limits_{a=1}^{a=\infty}\la_a'(\delta Y^a)^2. 
\label{dt2si}
\ee The $\la_a'$'s can be arranged in increasing order
$\la_1'\le\la_2'\le\la_3'\le \cdots$. Clearly 
the integral in (\ref{Om19}) will be convergent if $\la_1'>0$; $S$ is
then a maximum and
$\la_1'>0$ is   a    condition of stability. If
$N>>1$ and $\delta^2\si'$ is sufficiently small, 
$S_g\simeq S$ as is usually expected from a steepest descent
calculation. We can  in principle evaluate the limit of
validity of this quasi-equality by calculating $\delta^2\si'$.


\vskip .5 cm  
{\it (v)  Convergence of $\Om(E)$ in meanfield theory.}
\vskip .5 cm

There are two factors that  cause $\Om(E)$ to be zero if it converges:
$\bb$ as we said diverges and the spectrum of eigenvalues $\la_a'$ is
unbounded if the domain of definition of eigenfunctions is  bounded.
This is  true under very general conditions that do not depend on the
form of a finite domain \cite{CH}. It is true in particular for
isothermal spheres
\cite{HK1} as shown in appendix B. In other volumes we
may restrict  $\delta W$'s to functions that satisfy a Poisson
equation
$\Delta \delta \rho = 4\pi G\delta W$  and use these
$\delta\rho$'s as variables; they are zero outside  $V$. In that case,
either
$\delta^2\si'>0$, the Boltzmann entropy is not a maximum of $\si$
which tends to plus infinity and
$\Om(E)$ is undefined    in this approximation  because $\bb$ in the
denominator of (\ref{Om18}) divergence. Or
$\delta^2\si'<0$  and  the Gibbs density of states function  tends to
zero! There is no way out of this difficulty in a meanfield steepest
descent calculation\footnote{see also the end of Appendix B in this
respect.}.

The culprit of the trouble is easily traced to the Laplacian operator
whose inverse is the divergent $1/r$ potential
\cite{HK1}. Some form of  short distance cutoff should have the effect
of replacing the Laplacian by a non local operator with a convergent
set of negative eigenvalues $\la'_a$ and a convergent $\bb$. 

Notice one important point: that the  number of
nodes of the eigenfunctions  grows steadily as
$\la_a'$ increases. When the spacing between the
nodes in the finite domain is of the order of the short distance
cutoff  the Laplacian in $\delta^2\si'$ is no more a reasonable
representation for a convergent
$\Om(E)$. This means however that a short distance cutoff will changes
the higher end of the spectrum and that  $\la_1'>0$ is a perfectly
good stability condition for the meanfield configurations with a
Boltzmann distribution.  
\vskip .5 cm


{\it (vi) Local and global maxima.}
\vskip .5 cm
 
 Had we made a correct calculation with some specific cutoff not too
big, we would most likely have found at least two maxima of $\si$, the
one we have just found and another one associated with a highly
concentrated core and entropy $S_0$ say. This $S_0$ would tend to
infinity if the cutoff went to zero. A steepest descent calculation
would have given
\be
\Om(E)=e^{S_g}\simeq e^{S_0+S}
\ee
Since both $S$ and $S_0$ are of order $N>>1$ the Gibbs entropy, 
for most values of the energy $E$ and  volume $V$ would be
approximately equal to either $S$ or $S_0$. Such a transition appears
in a Monte-Carlo calculation of $\Om(E)$ by de Vega and Sanchez
\cite{dVS}.

In principle the global
maximum is the equilibrium configuration after an infinite time.
However, the system may find itself in the local maximum for a very
long time indeed. This makes local maxima interesting.


\vskip .5 cm
\noindent\normalsize{\it (vii) Remarks about relativistic systems and
other ensembles.}
\vskip .5 cm
 (a) Relativistic systems.
\vskip .5 cm
 The Gibbs density of states function can and has been calculated in
general relativity \cite{KH}. The  advantage of general relativity is
that it is a mean field theory to start with. The  disadvantage is
that the formalism is far more complicated,  arbitrariness of
coordinates creates additional difficulties in functional integrations
and moreover a density of energy is not  defined.  With so much
trouble, one finds however that the total  entropy  is again given by
(\ref{S}) but of course
$f_e$ is different from (\ref{f}).

Other ensembles have not found   wide applications in astronomy but
they are of theoretical interest.  
\vskip .5 cm 
(b) The grand canonical ensemble. 
\vskip .5 cm
 In this ensemble $N$ and $E$ are not kept fixed. What is
fixed is  $\al$ and $\beta$. It follows that the Gibbs
equivalent for this ensemble $\Om_{GC}$,  called the grand
partition function, is  in the meanfield approximation  given by
expression (\ref{Om4}) in which
${1\over 2\pi i}e^{-\al N}d\al$ and
${1\over 2\pi i}e^{\bt E}d\beta$ integrations are removed from $e^\si$
because $N$ and $E$ are not fixed anymore; thus
\be
\Om_{GC}={1\over\bb}\int_{-\infty}^{+\infty} e^ {\si_{GC}}{\cal
D}W~~\Rightarrow~~\si_{GC}=\si+\al N-\bt E.
\label{GCE}
\ee
$\Om_{GC}$ is thus a double Laplace transform of $\Om(E)$.
Thermodynamic equilibrium is associated with the extremum $\si_{GCe}$
of the exponent
$\si_{GC}$. This extremum is simply related to the entropy: 
\be   \si_{GCe}=S+\al N-\beta E.
\label{wGC}
\ee
 The thermodynamic potential has no conventional name. $N$ and $E$
are  now mean
 values  which are however related to 
$\al,\bt$ by the same equations (\ref{NE}). In this ensemble  like in
the following ones equations (\ref{Poisson}), (\ref{f}) and
 (\ref{NE}) always hold at the point where the exponent  is extremum.
But equations (\ref{NE}) have slightly different interpretations. 
\vskip .5 cm (c) The canonical ensemble.
\vskip .5 cm In this ensemble the system has a fixed number of
particles and the temperature or $\bt$ is kept fixed rather than the
total energy. The Gibbs equivalent for this ensemble $\Om_C$ is called
the partition function and it is obtained from (\ref{Om4}) by removing
the ${1\over 2\pi i}e^{\bt E}d\bt$ integration from $e^\si$. Thus
\be
\Om_C={1\over 2\pi
i\bb}\int_{a-i\infty}^{a+i\infty}\int_{-\infty}^{+\infty}e^
{\si_C}d\al{\cal D}W
~~\Rightarrow~~
\si_C=\si-\bt E .
\label{CE}
\ee The corresponding thermodynamic equilibrium function which plays
the role of $S$ and associated with the extremum $\si_{Ce}$ of the
exponent
$\si_C$ is 
\be \si_{Ce}=S-\bt E=-\bt F.
\label{w''}
\ee The thermodynamic potential $F$ is known as the free energy.

\vskip .5 cm (d) The grand microcanonical ensemble.
\vskip .5 cm 
This fourth ensemble  not often considered \cite{LK} is
one  in which the total energy is fixed but the particle number is
allowed to fluctuate. This ensemble is physically plausible if there
are pair creations of particles.  Thus
$\al$ and the total energy are fixed. The Gibbs equivalent $\Om_{GMC}$
is  obtained by removing the
${1\over 2\pi i}e^{-\al N}d\al$ integration from $e^\si$:
\be
\Om_{GMC}={1\over 2\pi
i\bb}\int_{b-i\infty}^{b+i\infty}\int_{-\infty}^{+\infty}e^
{\si_{GMC}}d\bt{\cal D}W
~~\Rightarrow~~
\si_{GMC}=\si+\al N .
\label{GMCE}
\ee
Other ensembles have been studied like isothermal spheres
under  constant pressure.


\vskip .5 cm
\noindent\normalsize{\it (viii) Summary and conclusions.}
\vskip .5 cm

Here now are the main points
of this section.

 \nnn  (1) We started from the Gibbs density of states function $\Om$,
equation (\ref{Om1}), which we wrote  in a meanfield approximation as
equations (\ref{Om4}) (\ref{si}).

\nnn (2) We used a steepest descent method to calculate equilibrium
configurations which are given by Poisson's equation (\ref{Poisson})
with a Boltzmann distribution of  particles in phase space  (\ref{f}).
We also calculated the entropy  (\ref{S}) and got
a meaningful expression if (\ref{Om18}) is
convergent. The condition of convergence is  a condition of
stability. The results are only valid if some sort of short distance
cutoff is assumed, the details of which are however unimportant. The
limit of validity of the whole approximation scheme can be evaluated
if $V$ is specified. It is not clear however if the Boltzmann entropy
is a global or a local maximum. That depends on the cutoff and it would
show up in a correct calculation.

Lynden-Bell and Wood, in dealing with isothermal spheres proceeded in
a  simpler way.  They started  from the  entropy defined by
 (\ref{S}) with an unknown distribution function $f$   restricted by
conditions (\ref{NE}) in which
$\bt^{-1/2}W=U$ is a solution of Poisson's equation  (\ref{Poisson})
with 
$\rho=m\int fd\om$.  This out of equilibrium entropy is then
extremized ($\delta S=0$) with arbitrary $\delta f$'s subject to two
restrictions 
$\delta N=\delta E=0$. The extremum of $f$  is the 
  Boltzmann distribution  function (\ref{f}). Padmanabhan \cite{Pad}
used from the same method to find stability conditions. This
method is simpler and well adapted to gaseous systems. It
short-circuits many of the difficulties encountered above. Thus the
statistical thermodynamics of Gibbs ends up with the same mathematics
but it must be said again that ``it has a particular beauty of its own [and]
is applicable quite generally to every physical system" \cite{Sch}. 

Section 1 dealt with the  thermodynamic equilibrium of  $N$ point
particles of mass $m$ isolated in a volume $V$ with
 energy  $E$ and  entropy $S$.  Section 2 deals with thermodynamic
functions which give statistical thermodynamics its physical content.
We also develop the thermodynamic  stability theory of equilibrium
configurations. 
The theory of stability  in section 2  applies
 to a much wider class of systems than the one studied in this section.


\sect{Thermodynamic equilibrium and stability }
 
The first subsection deals with the thermodynamics of the
system of
 $N$ particles with gravitational interactions in a volume $V$. This
is our generic case. The rest of section 2 is more general. {\it
Please notice} that we shall be concerned by equilibrium configurations
only; we drop therefore the index $_e$ from $\al_e, \bt_e$ and  $f_e$
which is we have no reason to use anymore and write simply
$\al, \bt$ and $f$. 


\vskip .5 cm
\noindent\normalsize{\it (i) Thermodynamic functions of self
gravitating particles.}
\vskip .5 cm

Besides $E,N,V$ and
$S(E,N,V)$ there are  other thermodynamic functions that give a
physical content to the results. Consider in particular the
derivatives of
$S$ with respect to
$E,N,V$. The derivatives are easily calculated by considering $\si$
defined in (\ref{si}). This is a function of $E,N,V$ as well as of
$\al,\bt,W({\bf r})$. But the entropy  is an extremum of $\sigma$ for
which
$\di \sigma/\di \al=\di \sigma/\di \bt=\delta \sigma/\delta W({\bf
r})=0$. Therefore the partial derivatives of S with respect  to
$E,N,V$ are the same as the partial derivatives of  $\si$ with respect
to $E,N,V$ keeping  $\al,\bt,W({\bf r})$ fixed. The partial
derivatives must of course be evaluated at the extremum. A quick look
at  (\ref{si}) shows immediately that: 
\be {\di S\over \di  E}=\beta   ~~~~,~~~~{\di S\over \di
N}=-\al~~~~,~~~~{\di S\over \di
V}=\left(\int_{-\infty}^{+\infty}fd^3p\right)_b={\rho_b\over m};
\label{TF}
\ee the index $_b$ means  ``on the boundary"\footnote{ There would be
no mechanical  equilibrium otherwise as the boundary must be an
equipotential. Indeed,
$\rho_b$ must be the same everywhere  on the boundary as  mechanical
equilibrium requires: $-\nabla P-\rho\nabla U=0$ which implies,  see
(\ref{P}), 
$U\propto\ln \rho$ and $U_e$ must be the same everywhere on the
boundary because  the pressure force is normal to the surface.}. It is
worth noticing that the local pressure 
\be 
P({\bf r})= \int_{-\infty}^{+\infty}{1\over 3}m\dot{\bf
r}^2fd^3p={1\over
\bt}\int_{-\infty}^{+\infty}fd^3p={\rho({\bf r})\over m\bt};~~~{\rm
thus}~~~{\di S\over \di V}=\bt P_b,
\label{P}
\ee 
where $P_b$  is the pressure on the boundary. It follows  from
(\ref{TF}) and (\ref{P}) that
\be dS=\bt dE-\al  dN+\bt P_bdV~~~{\rm or}~~~{1\over \bt}dS=dE-{\al
\over \bt}dN+P_bdV.
\label{firstlaw}
\ee This differential expression, the most important one in
statistical thermodynamics,  shows that
\be {1\over \bt}=T~ {\rm is ~the ~temperature,}~~~{\rm and
}~~{\al\over\bt}=\mu~{\rm is ~the ~Gibbs~chemical~potential}.
\ee
$T,\mu.P_b$ are global thermodynamic quantities; they concern the
whole  system.  


\vskip .5 cm
\noindent\normalsize{\it (ii) Conjugate thermodynamic functions.}
\vskip .5 cm

Of particular interest in stability analysis are pairs of conjugate
thermodynamic functions with respect to the entropy like
$(E,\bt=\di S/\di E)$, $(N,-\al=\di S/\di N)$ and
$(V,\bt P_b=\di S/\di V)$.  Pairs of conjugate  parameters in
other ensembles are easily obtained from (\ref{firstlaw}). For instance in
the grand canonical ensemble the pairs of conjugate parameters with
respect to $\si_{GCe}$ are $(\beta,-E)$,
$(\al,N)$ and $(V,\bt P_b)$ since from (\ref{firstlaw}) and
(\ref{wGC})  we  see that
\be d\si_{GCe}=-E d\bt+N d\al+\bt P_bdV.
\ee
In the canonical ensemble pairs of conjugate parameters associated
with the free energy~ or rather
$-\bt F$ are   $(\bt, -E)$ and $(N,-\al)$ and $(V,\bt P_b)$ and for
the grand microcanonical ensemble the pairs of conjugate thermodynamic
functions with respect to $\si_{GMCe}$ are $(E,\bt), (\al,N), 
\\(V,\bt P_b)$. From here on the thermodynamics of self-gravitating
systems can be developed along the royal path taken by Lynden-Bell and
Wood
\cite{LBW} for  studying isothermal spheres.
Now we continue at a more general level.


\vskip .5 cm
\normalsize

\nnn {\it (iii) Stability conditions for equilibrium states in
general.}
\vskip .5 cm  If the Gibbs density of states function in the mean field
approximation is convergent and  the steepest descent evaluation of
$\Om(E)$ to order two is meaningful, equation (\ref{Om18}) together
with equation (\ref{dt2si}) suggests that with a proper choice of
variables in 
$\Om(E)$, say
$x^a$, in  (\ref{Om4}) the stability conditions would appear
directly in their simplest form. Thus starting from {\it any} Gibbs
density of states function of  {\it any} ensemble  in these appropriate
variables we would write $\Om(E)$ in a form like this:
\be
\Om(E)= {1\oo\cc}\int e^{w(E;x^a)}\prod\limits_a dx^a,
\label{Om6}
\ee  where $\cc$ is a constant that depends on the choice of the
$x^a$'s. The extremum of $w$ which defines equilibrium configurations
is then of the form
\be {\di w\over \di x^a}=0 ~~~{\rm~whose ~solutions}~~~x^a=X^a(E);
\label{equil}
\ee there may be more than one solution for  given $E$ and whatever
other parameters there may be, say
$\zeta_q, (q=1,2,...)$. Near an extremum where $x^a=X^a(E)$, the
expansion of
$w$ would  be of the following form:
\be   w=w_e-{1\over 2}\sum\limits_a\la_a[x^a-X^a(E)]^2+O_3.
\label{stab2}
\ee Equations (\ref{equil}) and  (\ref{stab2}) are similar to those
obtained for  testing equilibrium and stability of a mechanical system
with potential energy $-w$, variables $x^a$ and parameters 
$(E,\zeta_q)$ though the number of variables in this case is  usually
finite. The whole analysis as can be seen is in fact
 valid for self-gravitating systems or other systems with short range
forces in thermal equilibrium.

The calculation of the stability limit $\la_1$ is  not  easy in
general. Therefore the following criteria may be helpful.


\vskip .5 cm
\nnn {\it (iv) A general criterion of stability. }
\vskip .5 cm

(a) A useful identity
\vskip .5 cm For definiteness we keep in mind the microcanonical
ensemble and the $w(=\si)$ introduced in the previous section but keep
our new notations that refer to more general systems.  Consider the
parameter\footnote{Any other parameter might do.}
$E$ and introduce its conjugate with respect to $w$ rather than with
respect to $w_e$. This defines a sort of inverse ``temperature out of
equilibrium" say
$\td\bt(E;x^a)$. The derivative of  $w$ given by equation
(\ref{stab2})  with respect to $E$ keeping all $x^a$'s and 
$\zeta_q$'s fixed:
\be
\td \bt(E;x^a)={\di w\over \di E}=\bt (E)+\sum\limits_a
\lambda_a{dX^a\over dE}[x^a-X^a(E)]+{\cal O}_2~~~,~~~\bt (E)={\di
S\over \di E}.
\label{tdbt1}
\ee Clearly the equilibrium value of $\td\bt$ is $\bt$ and the second
order terms is to be interpreted as fluctuations of $\bt$ near equilibrium.
When
$w=\si$,
$\bt$ is the inverse temperature; here $\bt$ is some ``generalized"
inverse temperature. 

Let us calculate first order derivatives of $\td\bt$ with respect to
$x^a$ and $E$  at the point of extremum. At that point  derivatives of
${\cal O}_2$ in (\ref{tdbt1}) are  at least  of order 1 and will thus
add nothing to the derivatives at the point of extremum. The second
terms of (\ref{tdbt1})  contribute  only through derivatives of
$[x^a-X^a(E)]$ because this quantity  is  zero at the extremum. One
thus obtain:
\be
\left({\di\td\bt\over\di x^a}\right)_e=\la_a\left({\di X^a\over \di
E}\right)~~{\rm no~summation~on~{\it a}}~,~~\left({\di\td\bt\over\di
E}\right)_e={\di\bt\over
\di E}-\sum\limits_a\la_a\lll{\di X^a\over \di E}\rrr^2.
\label{tdbt2}
\ee Extracting $\di X^a/\di E$ from the first equality and inserting
into the second gives  the following expression for the slope of the
``linear series"  of conjugate parameters $\bt(E)$ with respect to
$w_e$:
\be
\displaystyle{\di\bt\over \di  E}=\lll{\di\tilde\beta\over \di
E}\rrr_e+\sum\limits_a{(\di\tilde\beta /\di x^a)_e^2\over\lambda_a}~.
\label{slope1}
\ee

Let us emphasize again that the  calculation   is valid for any pair of
conjugate parameters with respect to any $w$, $e^w$ being the
statistical weight of any thermodynamic system, including one with
short range forces between the particles.  It also applies to any pair
of conjugate parameters with respect to (minus) the potential energy
of a mechanical system in equilibrium. It is however extremely useful
to keep in mind the concrete example of the $N$ gravitating particles
of mass
$m$ with energy $E$ in a volume
$V$.
\vskip .5 cm

(b) Poincar\'e's criteria of stability.
\vskip .5 cm There exists a classical result of Poincar\'e \cite{PO}
which is discussed in  various treatises on stability, see for
instance Jeans \cite{J} Ledoux \cite{LE} or Lyttleton \cite{LY},  and
was applied to self-gravitating systems in thermal equilibrium for the
first time by Lynden-Bell and Wood \cite{LBW}. It
 says the following.  A change of stability {\it may  only occur}
where two or more linear  series of equilibria like $\bt(E)$ for
instance\footnote{Any linear series, not only of conjugate
thermodynamic functions like $\bt(E)$. For instance, one of the 
$X^a$'s as a function of one of the $\zeta_q$'s or a $P(V)$ diagram.} 
have one equilibrium configuration in common (``bifurcation points")
or where two or more series merge into each other (``turning points").
When this happens stable equilibria {\it may} turn into unstable ones;
reciprocally, unstable ones may become either stable or more unstable.
Thus a change of sign of  $\la_1$ and perhaps of more eigenvalues or 
stability coefficients appear only at bifurcations and turning points. 

The Poincar\'e method of linear series does not say  whether a change
of stability actually occurs or not. It has been  pointed out
\cite{K78} that under very general conditions  the stability  or the
number of unstable modes (the number of negative $\la_a$'s) can be
deduced from the topological properties of series of equilibria, i.e.
from purely thermodynamic considerations as we shall now review. 
\vskip .5 cm
(c) Simple eigenvalues.
\vskip .5 cm Suppose the spectrum of eigenvalues is simple that is
non-degenerate:
$\la_1<\la_2<\la_3....$.   This assumption is far more than we shall
ever need but it makes further explanations simpler. In practice only
some of the smaller $\la_a$'s need to be distinct.  Consider   a   line
$\beta (E)$ with points  that correspond to stable configurations
($\la_1>0$) and follow the line towards a limit of stability.  As we
approach that limit 
$\la_1\rightarrow 0$  the
$1/\la_1$ term in (\ref{slope1}) begins to dominate and
\be {\di \bt\over \di E} \simeq{(\di\tilde\beta /\di
x^1)_e^2\over\lambda_1}\rightarrow+\infty~~~{\rm
provided}~~~(\di\tilde\beta /\di x^1)_e\neq 0.
\label{slope2}
\ee We shall come back later to what happens if  $(\di\tilde\beta /\di
x^1)_e\rightarrow 0$. If equation (\ref{slope2}) holds, the slope of
$\bt(E)$ has the same sign as $\la_1$. As a result:

(1) Two linear series of conjugate thermodynamic functions merge into
each other at a point where there is a vertical tangent. A change of
stability occurs thus necessarily there, at a turning point.

(2) The stable branch is the one with a positive slope near the
vertical tangent. Another way to state this property is as follows.
Stability is lost where the branch turns counterclockwise. A stable
branch never turns clockwise.

(3) Say $\la_1<0$ and $\la_2\rightarrow 0$ along the $\bt (E)$ line. 
There is thus already  one mode of instability. If a further mode of
instability shows up it must be at a new turning point where
$\la_2\rightarrow 0$. Near the new turning point the dominant
contribution to (\ref{slope1}) comes from the $1/\la_2$ term and
equation (\ref{slope2}) holds with an indice 2 instead of an indice 1.

(4) Say $\la_1<0<\la_2$ and  $\bt(E)$ turns clockwise. This means that
$\la_1\rightarrow 0$ again and that  unstable configurations  turn
into stable   ones on the side with positive slopes.

(5) If a stable linear series spirals inwards counterclockwise,
stability is lost at the first turning point. We give an example of
this case in section 4.

(6) If a change of stability occurs at a point where $(\di\tilde\beta
/\di x^1)_e\rightarrow 0$ and $\la_1\rightarrow 0$ the previous
considerations become obviously invalid. Notice however that  a  small
perturbation of  the potential energy  easily removes the
coincidence.   Indeed, let us change  $w$ to
$w+\epsilon Ex^1$ say with $|\epsilon|<<1$. The equilibrium equations
(\ref{equil}) become 
\be {\di w\over \di x^1}+\epsilon E=0 ~~(|\epsilon|<<1) ~~~{\rm and}
~~~{\di w\over \di x^a}=0 ~~(a>1).
\label{+epsilon}
\ee while
\be
\lll{\di\tilde \beta \oo\di x^1}\rrr_e=\lll{\di^2[w+\epsilon Ex^1]
\over\di E \di x^1}\rrr_e\stackrel{\la_1\rightarrow
0}\longrightarrow\epsilon \neq 0~~\Rightarrow~~{\di\bt\oo\di
E}\rightarrow {\epsilon^2\oo\la_1}.
\ee This shows that such coincidences will show up in the
mathematics but not
in 
 experimental observations  \cite{TH} nor are they likely to appear in
numerical calculations and many interesting  models in astronomy and
astrophysics are solved numerically. 

(7) Finally if we know that one configuration is stable or we know its
degree of instability we shall know about the stability or degree of
instability of every other configuration of the linear series. 
\vskip .5 cm
 (d) Bifurcations  and degenerate spectra.
 \vskip .5 cm Bifurcations like degeneracies of the spectrum of
stability coefficients   come often from an excess of symmetry. A
characteristic property of  bifurcations in problems of physical
interest is that they are ``unstable". Small perturbations of the
potential energy as just indicated remove unstable bifurcations and
transform them into turning points \cite{TH}; such  bifurcations  do
not show  in experiments or in numerical solutions. We shall give an
example of an unstable bifurcation in  section 4.   The theory of
bifurcations\footnote{The  problem of  distinguishing stable from
unstable  bifurcation offers little difficulty  with linear series depending
on one parameter 
\cite{AR}. The unfloding of bifurcations with more than one parameter
and several variables is  much more complicated. Thom \cite{Th} classified
all structurally stable inequivalent unfoldings or as it is called
``elementary catastrophes" with up to two variables and four parameters.
Great progress was done in this subject since  1975. 

Catastrophe theory
has not been very useful in physics. It did not contribute a thing that was
not already known otherwise. It provided however a sound mathematical
basis to the theory of  bifurcations and to the classification of
inequivalent ones. } is of some interest in mathematical physics \cite{AR}. 

 In the rest of this section we  deal only  with simple spectra and
turning points at vertical tangents. 


\vskip  .5 cm
\normalsize

\nnn {\it (v) $C_V<0$ and the nonequivalence of ensembles.}
\vskip  .5 cm

(a) About $C_V<0$.
\vskip  .5 cm We can now see that in a thermally stable isolated system
but {\it near instability}
$(\la_1\rightarrow 0)$ the heat capacity
$C_V=\di E/\di T$ is negative in stable configurations and turns
positive in unstable configurations by going through zero. Indeed, 
according to (\ref{slope2}),
\be C_V={\di E\over \di T}=-\bt^2{\di E\over \di\bt}\simeq
-\la_1\bt^2\left({\di\tilde\bt \over\di x^1}\right)_e^{-2}<0.
\label{CV1}
\ee The same will of course happen in an unstable system on the verge
of acquiring a second mode of instability
$(\la_1<0<\la_2\rightarrow 0)$.   $C_V$ may thus also {\it be negative
in unstable systems}. The sign of $C_V$ 
 is therefore not a criterion of stability like in classical
thermodynamics.

The fundamental reason for this difference is the fact that the system
is no subsystem of an ensemble. This is made obvious by the following
standard explanation of why small subsystems are stable if their
heat capacity is positive. 

Consider an isolated system of energy $E$ made up of two systems of
energy $E_1$ and $E_2=E-E_1$ of comparable size.  Let
 $S_1(E_1)$ and $S_2(E_2)$ be their entropies. The sum of the
entropies is the entropy of the system {\it out} of equilibrium but
for simplicity we write  
\be S(E)=S_1(E_1)+S_2(E_2).
\ee Equilibrium is reached when
\be
\delta S=\left[\left({dS_1\over dE_1}\right)_e-
\left({dS_2\over dE_2}\right)_e\right]\delta
E_1=(\beta_1-\beta_2)\delta E_1=0~~~\Rightarrow~~~\beta_1=\beta_2\equiv
\beta.
\ee To check stability we  calculate second order variations of $S$
which at equilibrium must be negative:
\be (\delta^2S)_e={1\over 2}\left[\left({d\beta_1\over
dE_1}\right)_e+\left({d\beta_2\over dE_2}\right)_e \right](\delta
E_1)^2=-{1\over 2}\bt^2\lll{1\over C_{V1}}+{1\over C_{V2}}\rrr(\delta
E_1)^2<0.
\ee 
 Since $\delta E_1=C_{V1}\delta T_{1}=-C_{V1}\delta \bt_1/\bt^2$, we
can rewrite the last expression as follows:
\be
\lll\delta^2S\rrr_e=-{1\over 2}\lll 1+{C_{V1}\over C_{V2}}\rrr C_{V1}
\lll {d\beta_1\over
\beta}\rrr^2<0.
\ee 
Now if system 1 is a small subsystem, $E_1\ll E_2$, also
$C_{V1}\ll C_{V2}$ and
\be (\delta^2S)_e\simeq -{1\over 2}C_{V1}\lll{\delta\beta_1\over \beta
}\rrr^2<0.
\label{dt2S}
\ee The stability conditions (\ref{dt2S}) implies that a stable  small
sub-system must have  a positive heat capacity as is well known. 
\vskip  .5 cm
 (b) About the non-equivalence of ensembles.
\vskip  .5 cm
 If instead of keeping the system isolated we keep it at
constant temperature, the role of $S$ is then  played by the
thermodynamic potential $w_{Ce}$ or $-\bt F$, see equation (\ref{w''}),
and $(\bt ,-E)$ is  a pair of conjugate parameters with respect to
$-\bt F$. But $-E(\bt)$ is the same line as $\bt(E)$ drawn in
coordinates rotated $90^\circ$ clockwise. Vertical tangents become
horizontal ones and reciprocally. Let $\la_{aC}$ be the Poincar\'e
coefficients of stability for the canonical ensemble. Near instability
$\la_{1C}\rightarrow 0$. The analog of (\ref{slope2}) is now 
\be {\di (- E)\over \di\bt}\simeq{\LLL\di (- \td E) /\di
x^{1}_C\RRR_e^2\over\lambda_{1C}}>0~~~\Rightarrow~~~C_V={\di E\over \di
T}=\bt^2{\di (-E)\over
\di\bt}>0.
\label{slope3}
\ee
 This shows  that near  instability a stable canonical ensemble has 
$C_V>0$. Nevertheless unstable configurations may also have $C_V>0$.
For instance if $\la_{1C}<0<\la_{2C}$ and $\la_{2C}\rightarrow 0$. Thus
$C_V>0$ is not a criterion of stability for a canonical ensemble either.  It is
surely  a sufficient condition of stability  near  and indeed far away from
instability.

The canonical ensemble and the isolated system or microcanonical
ensemble cannot become unstable for the same equilibrium
configuration  unless vertical and horizontal tangents appear
at the same point. This difference between ensembles is
referred to as the non-equivalence of ensembles and is a property of
non-extensive systems like self-gravitating ones. In extensive
systems, different ensembles are equivalent \cite{LL}.
Why does
the canonical ensemble get unstable and the other does not is nicely
explained in the Lynden-Bell and Wood paper \cite{LBW}.


\vskip  .5 cm
\normalsize

\nnn {\it (vi) Summary and conclusions.}
\vskip  .5 cm 

Under general conditions, mainly that the spectrum of
Poincar\'e coefficients of stability be simple, stability limits and
the number of unstable modes can be found for all calculated
configurations using only linear series of thermodynamic functions of
equilibrium configurations provided we know whether {\it one}
configuration is stable or what is its degree of instability. Conjugate
thermodynamic functions are particularly valuable in this respect as
changes of stability show up at  vertical tangents and the slopes in
their vicinity have simple interpretations.

The thermodynamic criterion has a number of limitations. One rarely
calculate all the sequences of equilibrium and therefore some bifurcations
may not show up because the branch points are missing. Thus equilibria
might become unstable and the system might choose to be in a more
stable state which has not been calculated. A complete answer needs a
detailed analysis of the second order terms of the Gibbs density of
states function like the one started in subsection $2(i)$. The much
simpler version of Padmanabhan \cite{Pad89} is also useful. 

A less severe limitation is that the method is mathematically not a
full proof because we must assume that the spectrum of stability
coefficients is simple and at the same time that conditions like
(\ref{slope2}) hold. The method may however be applied to numerical
solutions with a reasonable degree of confidence because this type of
mathematical singularity
  will not show up in numerical calculations nor indeed in
experiments. 

Another general  limitation of  any thermodynamic criterion of
stability  is that we learn little about the nature of 
instabilities, triggering mechanisms, and what becomes of stable states
which evolve through a series of quasi-equilibria along the linear  series up
to and beyond the limit of instability.

We have seen that the sign of heat capacities is not a criterion of
stability and that in microcanonical and canonical ensembles near
instability heat capacities have opposite signs. These ensembles are
not equivalent as in classical thermodynamics. The proofs of these
properties have a great degree of generality.
The next section deals with fluctuations.

\sect{Fluctuations}

\normalsize

\nnn {\it (i) Fluctuations in self-gravitating systems.}
\vskip  .5 cm 
The thermodynamics of fluctuations in stable equilibrium
configurations becomes interesting near instability. Fluctuations can
put the system out of equilibrium.  Calculations of fluctuations are
particularly interesting when dealing with metastable states of
systems like  the system of point particles in an enclosure considered
in section 1. The entropy of such a system may be a ``local" maximum 
smaller than the  absolute maximum. In that case there are  two
maxima and between them there must  a saddle point. Such a saddle point
  must    show up close to a limit of stability because there
the linear series turns counterclockwise. Figure 1 illustrates the
situation with a linear series
$\bt(E)$ for  isolated systems. Let
$\bt$ be the inverse temperature of a stable system and $\bt '<\bt$ the
inverse temperature of the unstable system with the same energy $E$.
If the mean quadratic fluctuation of temperature or rather of its
inverse
$<(\delta \bt)^2>$ is equal to or greater than the square of the gap
$(\bt-\bt')^2$, the statistical weight of a {\it real}
 fluctuation bigger than $(\bt-\bt')$ becomes important; this may  put
the system in a state where $w$ can increase to a bigger  local maximum
or to the global maximum. That is the entropy will have a  chance to
increase which it will surely do and equilibrium will be lost. It
is thus interesting to evaluate the statistical weight of the
fluctuations. Fluctuations  can only displace the limit of stability
  towards  higher values of $\bt$  and the question is  how much
higher. This is the subject of this section. The theory  which has the
same degree of generality as the theory of stability of section 2,
comes from  works  by  Okamoto, Parentani and myself
\cite{OKP}\cite{PKO}.  
\vskip  .5 cm
\normalsize

\nnn {\it (ii) The probability of a fluctuation near instability.}
\vskip  .5 cm 
Figure 1 is a good starting point. The point of marginal
stability has coordinates $(E_m,\bt_m)$ and a vertical tangent. 
Consider near that point a stable configuration  with coordinates
$(E,\bt)$. The dominant contribution to
$\Om$ near instability, see (\ref{Om6}),  must come from $\delta x^1$
in $w$ since
$\Om\rightarrow
\infty$ when
$\la_1\rightarrow 0$. Let us therefore integrate over all  variables 
except $x^1$:
\be
\Om(E)\simeq{1\oo\cc}\int_{-\infty}^{+\infty} e^ {S-{1\over
2}\sum\la_a(\delta x^a)^2}
\prod\limits_{a=1}^{a=\infty}dx^a\simeq
{e^S\oo\cc}\prod\limits_{a=2}^{a=\infty}\lll{2\pi\oo
\la_a }\rrr^\half\int_{-\infty}^{+\infty}e^ {-{1\over 2}\la_1(\delta
x^1)^2}d(\delta x^1).
\label{Om7}
\ee As long as the exponent is sufficiently steep (thus not too close
to where $\la_1=0$) we may integrate over $\delta x^1$ between
$\pm\infty$ as indicated. 

Notice that  $\delta x^1=x^1-X^1$ where $X^1$ is the equilibrium value
of
$x^1$ at point
$(E,\bt)$ and $x^1$ itself is some value out of equilibrium with the
same energy
$E$. We have a relation between
$\delta x^1$ and the fluctuation of temperature
$\delta\bt=\td\bt(x^1;E)-\bt$ which to first order is simply
\be
\delta\bt=\td\bt-\bt\simeq\lll{\di \tilde\beta\over \di
x^1}\rrr_e\delta x^1.
\label{dtbt}
\ee
 Replacing $(\di
\tilde\beta /\di x^1)_e$  in terms of $\di E/\di\bt$ using the second
of  equations (\ref{slope2}),
\be
\delta x^1
\simeq{\delta\bt\oo(\di\td\bt/\di x^1)_e}\simeq\pm\lll{1\oo\la_1}{\di
E\oo\di\bt}\rrr^{1/2}(\td\bt-\bt).
\ee We use both these  equalities to replace  $\delta x^1$  in the
integral of (\ref{Om7}), the first equality in
$d(\delta  x^1)$ the second in the exponent, and obtain
\be
\Om(E)\simeq {e^S\oo\cc(\di\td\bt/\di
x^1)_e}\prod\limits_{a=2}^{a=\infty}\lll{2\pi\oo
\la_a }\rrr^\half\int_{-\infty}^{+\infty} e^{-\half {\di
E\oo\di\bt}(\td\bt-\bt)^2}d(\td\bt-\bt).
\label{Om8}
\ee  We conclude that  the statistical weight  of a fluctuation of
$\bt$ to between $\td\bt$ and
$\td\bt+d\td\bt$ is proportional to the integrand of (\ref{Om8}). Thus
the probability
$d\pp$ of  the fluctuation is given by the normalized expression:
\be d{\cal P}={1\oo \sqrt{\pi}}e^{-t^2}dt ~~~{\rm with }~~~t^2=\half {\di
E\oo\di\bt}(\td\bt-\bt)^2.
\label{dP}
\ee The mean quadratic fluctuation of temperature is  thus given by
\be 
\int_{-\infty}^{+\infty}\lll
{\delta\bt\oo\bt}\rrr^2d{\cal
P}=\big<\lll{\delta\bt\oo\bt}\rrr^2\big>=\big<\lll{\delta T\oo
T}\rrr^2\big>={1\oo(-C_V)}>0.
\ee Here we used (\ref{slope3}) to get the expression on the right
hand side. The formula is like  that of Landau and Lifshitz \cite{LL}
for  mean  quadratic  fluctuations of temperature {\it in a small 
subsystem}; it has by necessity the opposite sign for $C_V$. 

 Fluctuations of other mean values {\it near instability}  in different
ensembles can be derived in  a similar way.

The  formula (\ref{dP}) may be made more specific because we
are near the turning point where $(\di E/\di\bt)_m=0$. $(\di
E/\di\bt)$ is thus small and to the lowest order in
$(\bt-\bt_m)$, assuming  $(\di^2E/\di\bt^2)_m\ne 0$, 
\be {\di E\oo\di\bt}\simeq \lll{\di
^2E\oo\di\bt^2}\rrr_m(\bt-\bt_m)=\lll\bt^3{\di
^2E\oo\di\bt^2}\rrr_m\lll{\bt-\bt_m\oo\bt_m}\rrr\bt_m^{-2}=\lll T{\di
C_V\oo\di T}\rrr_m\lll{\bt-\bt_m\oo\bt_m}\rrr\bt_m^{-2}.
\label{dEdbt}
\ee
 $t^2$  defined in (\ref{dP}) can also be written
\be t^2=\half \lll T{\di C_V\oo\di
T}\rrr_m\lll{\bt-\bt_m\oo\bt_m}\rrr\lll{\td\bt-\bt\oo\bt_m}\rrr^2,
\label{t^2}
\ee and the mean quadratic fluctuation satisfy the following equation:
\be
\lll T{\di C_V\oo\di
T}\rrr_m\lll{\bt-\bt_m\oo\bt_m}\rrr\big<\lll{\td\bt-\bt\oo\bt_m}\rrr^2\big>~=1.
\label{fluc}
\ee If $(\di^2E/\di\bt^2)_m= 0$ a higher even  derivative must be
different from zero because $E$ is a minimum and the final results 
will not be very different;  there will be higher powers of
$\bt-\bt_m$.
\vskip  .5 cm
\normalsize


\nnn {\it (iii) Stability limits induced by fluctuations.}
\vskip  .5 cm
 We now define a new limit of stability, the point
$(\bt_l,E_l)$ on the linear series in figure 1 where the mean
quadratic fluctuation equals the square of the difference between
$\bt$ and  $\bt'$ for reasons explained at the beginning of this
section. Near the point of marginal stability
$(\bt_m,E_m)$ we have approximately  
$(\bt_l-\bt_l')\simeq 2(\bt_l-\bt_m)$; thus $\bt_l$ is defined by the
condition that
\be <(\td\bt-\bt_l)^2>\simeq 4(\bt_l-\bt_m)^2.
\ee We shall obtain $\bt_l$ by inserting this value of the mean square
fluctuations into (\ref{fluc}); this gives
\be
\lll 4T{\di C_V\oo\di T}\rrr_m\lll {\bt_l-\bt_m\oo\bt_m}
\rrr^3=1~~~\Rightarrow~~~\bt_l=\bt_m\LLL 1+\lll 4T {\di
C_V\oo\di T}\rrr_m^{-1/3}\RRR~.
\label{btl}
\ee Notice that $C_V$ is of order $N$ and therefore the change from
$\bt_m$ to
$\bt_l$ is of order $N^{-1/3}$ which is not necessarily very small. The
corresponding value of  $E_l$ is readily found by expanding
$E(\bt)$ in a Taylor series to order two near $\bt_m$:
\be E_l \simeq E_m+ \half  \lll 
{\di^2E\oo\di\bt^2}\rrr_m(\bt_l-\bt_m)^2 ~~\Rightarrow
~~E_l=E_m+{1\oo 8}\lll 4T {\di C_V\oo\di T}
\rrr_m^{1/3}\bt^{-1}_m~.
\ee  

To have some idea on how sharply defined the point $(\bt_l,E_l)$ is,
consider  an equilibrium configuration with $\bt>\bt_l$ and evaluate
{\it the weight} of the probability for a fluctuation  to induce
instability, i.e.  calculate $e^{-t^2}$ for 
\be
\td\bt-\bt\simeq 2(\bt-\bt_m). 
\ee If we insert this value  for $\td\bt-\bt$ into  (\ref{t^2}),
taking account of  (\ref{btl}) we find that 
\be t^2=\half \lll  {\bt-\bt_m\oo \bt_l-\bt_m}\rrr^3.
\ee For $\bt-\bt_m=\bt_l-\bt_m$,  $e^{-t^2}\simeq 0.6$ but if
$\bt-\bt_m$ is twice as big as $\bt_l-\bt_m$,  $e^{-t^2}\simeq 0.02$
and   if equal to three times that difference, 
 $e^{-t^2}\simeq 10^{-6}$. The exponential fall is relatively sharp
and the new limit of stability $(E_l,\bt_l)$ induced by fluctuations
is not too badly
 localized.
\vskip  .5 cm
\normalsize

\nnn {\it (iv) Summary and comments.}
\vskip  .5 cm
 Fluctuations in non-extensive systems like
self-gravitating ones have a peculiar behavior which has no analogue
in classical thermodynamics where smooth turning points do not exist
and ensembles are  equivalent. Fluctuations can  induce instability at
lower temperatures (higher $\bt$'s)  than the
theoretical limit  $\bt_m$.  The relative change is of order
$N^{-1/3}$. $N=10^3$ is often considered in numerical
calculations to study globular cluster models. Instability limits
found in these models may be quite different from more realistic ones
with $N=10^5$ or higher. Moreover, since $N^{-1/3}$ is not necessarily
a big number the stability limit  may be  sensitive to the values of
$(4T\di C_V/\di T )_m$.

Fluctuation theory is useful  when slow evolution occurs towards
instability as we shall see   in  section 5.


\sect{Examples}
In this section we give two
examples in which thermodynamics of self-gravitating systems is used.
The first one is that of isothermal spheres. Equilibrium
configurations were obtained numerically. In this  case linear series
of conjugate thermodynamic functions correctly give stability limits.
The calculation of fluctuations  give useful additional information.
The second example is that of liquid ellipsoids.  Here equilibrium
configurations are defined  by quadratures and the multiple  linear
series are replete with bifurcations. The bifurcations are however
unstable and may be lifted with small perturbations of the potential
energy. Other examples are mentioned in less detail. In some of these
examples, the stability theory of section 3  has been used and in
others  it has not but might have saved  complicated calculations to
obtain the same results had it been used.      
\vskip  .5 cm
\normalsize


\nnn {\it (i) Isothermal spheres.}
\vskip  .5 cm Isothermal spheres belong to  standard literature in
astronomy \cite{BT}. This well posed mathematical problem can be
solved numerically and provides a not too crude model for  cores of 
globular clusters and played an important role in their understanding. 
\vskip  .5 cm
\normalsize

 (a) Equilibrium configurations.
\vskip  .5 cm The problem at hand is the one treated in section $2 {\it
(iii)}$  with a spherical enclosure of radius
$r_B$ say. Antonov \cite{AN} has given an elegant proof  that in a
sphere, spherical configurations maximize the entropy. Spherical
equilibria have a  gravitational potential $U$ which depends on the
distance to the center  $r$ only and Poisson's equation
(\ref{Poisson}) with a Boltzmann distribution (\ref{f}) reduces in
this case to 
\be {1\oo r^2}{d\oo dr}\LLL r^2{dU(r)\oo dr}\RRR=4\pi G\rho
(r)~~~\Rightarrow  ~~~
\rho(r) =m\int_{-\infty}^{+\infty}fd^3p=m\lll {2\pi
m\oo\bt}\rrr^{3/2}e^{\al-m\bt U}.
\label{U(r)}
\ee 
The boundary conditions for regular solutions are:
\be
 \lll{dU\oo dr }\rrr_{r=0}=0~~~{\rm and}~~~U(r_B)\equiv U_B=-{GNm\oo
r_B}={-GM\oo r_B}.
\label{UB}
\ee Equation (\ref{U(r)}) has a singular solution
\be
\rho_{sing}= {1\oo 2\pi Gm\bt r^2}.
\ee Regular solutions oscillate around this singular solution to which
they become closer and closer as $r$ grows.

Excellent tables of numerical solutions have been given by Emden
\cite{EM},  more detailed ones are found in   Chandrasekhar 
\cite{CH39} with some additions in Chandrasekhar and Wares
\cite{CW}. The tables are still useful today even if Mathematica  
provides  $10^2$ exact digits in a fraction of a second for solutions
with density contrasts up to $10^8$ in no time.

  The density contrast $\rr\equiv \rho (r=0)/\rho (r=r_B)$ is a useful
parameter which characterizes the relative strength of gravity. It is
obtained by direct integration of equation (\ref{U(r)}) because
$U\propto
\ln
\rho$. Equilibrium configurations have been studied in detail in
Chandrasekhar
\cite{CH39} but also in   Lynden-Bell and Wood \cite{LBW} and in
Padmanabhan
\cite{Pad}. A short review on isothermal spheres  relevant to astronomy
is given in Binney and Tremaine \cite{BT}.
\vskip  .5 cm
\normalsize

 (b) Stable configurations.

\vskip  .5 cm
\normalsize
 
\nnn ($\al$) Isolated spheres and
isothermal spheres in a heat bath.
\vskip  .5 cm 
Antonov \cite{AN} found analytically that stable
isolated systems can exist for density  contrasts\footnote{In
general we give  numerical results with 3 digits.} 
$\rr < 709$. A  nice and simple  analytical proof has been given  by
Padmanabhan
\cite{Pad89}. Lynden-Bell and Wood \cite{LBW} found the same result
using a combination of Poincar\'e's turning point method and  direct
calculation of the entropy. They also found that equilibrium
configurations maintained  at constant temperature are stable for
$\rr<32.1$. It is worth noting that Padmanabhan's method allows to
calculate the density perturbations   $\de\rho$ that trigger
instability. This gives interesting additional information that is
beyond  thermodynamic's capability.

The method of linear series with  conjugate thermodynamic functions
give  these results in one stroke and adds new interesting
information. Figure 2 represents $\bt m|U_B|$ as a function of
$E/M|U_B|$;  $N$, 
$r_B$ and thus $U_B$ are  constant. 

For small $\bt$'s, i.e.   high temperatures, gravity plays a minor
role, the particles behave like a perfect gas and  perfect gases  are
stable for arbitrary perturbations \cite{LL}. The curve of figure 2
spirals inwards counterclockwise towards a point with coordinates
$E/M|U_B|=-{1\oo 4},~
\bt m|U_B|=2$.  If we follow  the stable branch
 from low density contrasts, say from $E=0$ where $\rr\simeq 6.85$, to
high density contrasts and negative energies, a  vertical tangent
appears at  $\rr \simeq709$. Thus isolated isothermal spheres become
unstable for  $\rr >709$. This confirms once again Antonov's result
but  figure 2 also shows that {\it all} configurations with a  density
contrast
$\rr >709$  are unstable with respect to spherically symmetric
perturbations and that the number of unstable modes increases with
$\rr$ because the linear series spirals inwards counterclockwise.

Notice as expected that stable configurations near the vertical
tangent at $\rr\simeq 709$ have negative heat capacities.  Notice
however that  unstable configurations with
$5.22\cdot 10^3<\rr<4.50\cdot 10^4$ have also $C_V<0$ though they are
unstable and have one mode of instability: $\la_1<0<\la_2$. 

Notice finally that all configurations are stable for non-spherically
symmetric perturbations \cite{HK} but this has not and can not be
proven by  thermodynamic arguments since  we have   linear series of
spherically symmetric equilibria only.

If we turn figure 2 clockwise  $90^\circ$  we are looking at  
$-E(\bt)$. This is the appropriate linear series for  canonical
ensembles as we have seen in section 2.  We can follow again the line
of stable configurations from $E=0$ and $\rr=6.85$ towards higher
density contrasts  with negative energies,  we shall meet a vertical
tangent at
$\rr=32.1$. Thus  isothermal spheres in a heat bath become unstable
for  $\rr >32.1$. This confirms Lynden-Bell and Wood's result and
shows in addition that all configurations with higher  density
contrasts are not only unstable but more and more so as
$-E(\bt)$  spirals inwards counterclockwise. 

Notice as expected that stable configurations near the vertical tangent
have  $C_V>0$ but  unstable configurations with two modes of
instability
$\la_1<\la_2<0<\la_3$ and density contrasts
$4.50\cdot10^4<\rr<5.45\cdot 10^5$ have also $C_V>0$. 

Chavanis used a method similar to that of Padmanabhan to calculate
the  density perturbations  $\de\rho$ that trigger instability
in isothermal spheres at constant temperature \cite{Chav1} as well as  in
the ensembles studied in the next subsection \cite{Chav3}.
\vskip  .5 cm
\normalsize

\nnn ($\bt$) Grand canonical
ensembles  and grand microcanonical ensembles.  
\vskip  .5 cm

The method of linear series with  conjugate thermodynamic functions
has also been applied  to systems in which neither  energy nor the
number of particles are fixed. In  grand canonical ensembles $\al,
\bt,  r_B$ are fixed.  An appropriate linear series for this ensemble
is, see section 3, the curve
$-E(\bt)$ at fixed $\al$
 and $r_B$. This linear series, see also \cite{LK},  is  shown in
figure 3A and 3B and must be looked at  rotated 
$90^\circ$ clockwise. What is represented is actually
$\bt^*=\bt/\bt_0$ as a function of
$E^*=(\bt_0^2Gm^2/r_B)E$ with $\bt_0=2\pi^3m^7(8\pi
Gr_B^2e^\al)^2$. For
$E>0$ , the non-dimensional quantities are of comparable magnitude.
Notice that near
$E=0$ there is a cross over but not a bifurcation with two different
solutions whose  stability or instability is known.  For
$E<0$ the curve winds in counterclockwise staying very near the
vertical axis. The limit point has coordinates   
$(E^*=-5.37\cdot10^{-3},\bt^*= 13.6)$. This is why the diagram has a
different scale  for $E<0$, figure 3a, and $E>0$, figure 3b. 

At  low   density contrasts and high temperature the energy tends to
zero; kinetic and potential energies are both high and  of  comparable
magnitude. It has been shown
\cite{HK1}   by calculating fluctuations of $w_{GC}$ near the
extremum, see (\ref{wGC}), that grand canonical ensembles  are 
stable with respect to arbitrary perturbations\footnote{Paper
\cite{HK1} contains correct calculations but  the wrong
interpretation. The authors believed they were dealing with isolated
systems when they dealt in fact with a grand canonical ensemble.} for
$\rr<1.58$.  We can see on figure 3 that   grand canonical ensembles
have indeed a turning point at
$\rr=1.58$,
$-E(\bt)$ has a vertical tangent there and turns counterclockwise. At
the next vertical tangent $\rr\simeq 106$ and the curve turns again
counterclockwise   and at higher density contrasts the curve is
spiralling inwards. There is thus no stable equilibrium configuration
for $\rr>1.58$. 

The situation is more eventful in the grand microcanonical ensemble in
which $E,
\al, r_B$ are fixed. Here figure 3a and 3b are appropriate linear series for
checking stability.  At low density contrasts equilibrium
configurations are certainly stable:  the ensembles are more
constrained than the grand canonical one which is stable.   We  see 
that stable configurations exist for  density contrasts
$\rr<1.66$. Configurations with
$1.66<\rr<11.6$ have one mode of instability. There exists  a second
series of equilibrium configurations with
$11.6<\rr<92.6$. At higher density contrasts the curve spirals inwards
counterclockwise and   equilibrium configurations  become more and
more unstable.

\vskip  .5 cm

\normalsize

(c) Fluctuations. 
\vskip  .5 cm 
Fluctuations  displace the limits of stability towards
lower density contrasts. For instance in isolated isothermal spheres
it was found  \cite{KO} that quadratic fluctuations would induce
instability at density contrasts 
$\rr_l\simeq 709\cdot e^{-3.30\cdot N^{-1/3}}$. 
If $N=10^3$, a number close to numerical experiments, $\rr_l
\simeq 510$. The formula is still marginally valid for
$N= 10$ for which $\rr_l\simeq 154$. These are not small
effects. 

In this connection, it is worthwhile  recalling Monaghan's \cite{MO}
application  of the theory of {\it hydrodynamic} fluctuations to a
self-gravitating gas.  He showed  that density fluctuations become
large before the point of ordinary  stability is reached.
Thermodynamics supports Monaghan's finding.

\vskip  .5 cm
\normalsize

 (d) Relevance to astronomy. 
\vskip  .5 cm
 The simple model of isothermal spheres  played an important role in 
understanding the structure of globular clusters. The model taken
seriously by observers \cite{MH}   is that of Michie \cite{MI} which
was put to extensive use   by King
\cite{KI}. The Michie-King model is a truncated Boltzmann distribution
which in our notation is this, see (\ref{f}):
\be f=A(e^{\bt E_0}-e^{\bt E_c})~~~{\rm for}~~~E_0\le E_c~~~{\rm
and}~~~f=0~~~{\rm for}~~~E_0> E_c. 
\label{MK}
\ee
$E_c$ is an energy cutoff that simulates the absence of high energy
escapers. Stability limits for this model and   variants thereof
\cite{KHD}  do not change the general trend of equilibrium
configurations \cite{K80}. Linear series  are always counterclockwise
inwinding spirals and stable configurations exist up to some maximum
density contrast that varies not very much. 

The distribution (\ref{MK})  is a simple model for classifying
observable parameters of globular clusters.

It is not clear what ensemble represents best  thermal equilibrium in 
cores of globular clusters. However, that  a stability limit exists is
now commonly accepted.  Globular clusters in quasi-thermal equilibrium
evolve slowly towards greater entropy and density contrast due to
stellar evaporation \cite{BT}. They then reach a limit beyond which
they cannot  stay isothermal. The  great contribution of Antonov
\cite{AN} was to point out this instability.

What happens when a core becomes unstable was explained by Lynden-Bell
\cite{LB68}. The explanation is now part of standard texts  \cite
{BT}. Stars diffuse towards the center which becomes denser  and
denser while isothermal equilibrium is lost. This phenomena known as
the {\it gravothermal catastrophe}  leads eventually to core collapse.
This has been confirmed by numerous studies mentioned  in Meylan and
Heggie \cite{MH}. 

Observational evidence by various authors, see in particular Trager,
Dorjovsky and King's paper \cite {TDK},  came in the late 1980's when
CCD observations allowed a systematic investigation of the inner
surface brightness profile of globular clusters. These authors 
classified globular clusters into two different classes. About $80\%$
with a projected density profile that fitted Michie-King models and
the rest with a density profile corresponding to a singular density  in
$1/r^2$. These cores are considered to have collapsed.

The work of Antonov and Lynden-Bell is the single most important
contribution to observational astronomy based on   thermodynamics
of self-gravitating systems.

\vskip  .5 cm
\normalsize


\nnn {\it (ii) Maclaurin and Jacobi ellipsoids.}
\vskip  .5 cm
 Liquid ellipsoids like isothermal spheres belong to
standard literature. Chandrasekhar
\cite {Chandra69} devoted a whole book to the subject. Liquid
ellipsoids have few independent variables and plenty of bifurcations.
We give here one example  in which a bifurcation is lifted with a small
perturbation of the potential energy as discussed in Section 3.
Solutions of Poisson's equation of equilibrium which involve elliptic
integrals  are taken from Chandrasekhar's book. The stability analysis
presented here is taken from
\cite {K79}. 
\vskip  .5 cm
 (a) Equilibrium configurations. 
\vskip  .5 cm
 Consider a  self-gravitating ellipsoid with uniform
density $\rho$ and semi-axis
$a\ge b\ge c$. It rotates around the $c$-axis with uniform and
constant angular velocity $\Upsilon$. The total mass $M$ and angular
momentum $L$ are constant:
\be M= {4\pi\oo 3}\rho abc~~~,~~~L={1\oo 5}M(a^2+b^2)\Upsilon\equiv
I\Upsilon.
\ee The motion described in comoving coordinates along $a,b,c$ has an
effective potential energy $V$ which is, see for instance \cite{LE} :
\be V=\half\int \rho Ud^3x+{L^2\oo 2I}.
\ee
$U$ is the gravitational potential. There are three variables in $V$,
$a,b,c$, and one constraint $M$. There are thus two independent
variables, say,
\be x^1\equiv({a\oo c})^2\ge x^2\equiv({b\oo c})^2\ge 1.
\ee The explicit form of $V$ is given in Chandrasekhar;  in our
notations,
\ba -V&=&{3GM^2\oo 10}\lll{4\pi\rho\oo
3M}\rrr^{1/3}F({x^1,x^2;s})~~~{\rm with} ~~~s={25L^2\oo
3GM^3}\lll{4\pi\rho\oo 3M}\rrr^{1/3}~~~{\rm and}\nonumber\\ F&=&
(x^1x^2)^{1/6}
\int_0^\infty {d\nu\oo\sqrt{(1+\nu)(x^1+\nu)(x^2+\nu)}}-
s{(x^1x^2)^{1\oo 3}\oo x^1+x^2}.
\ea Equilibrium configurations $X^1(s)$ and  $X^2(s)$ are solutions of 
\be {\di V\oo\di x^1}={\di V\oo\di x^2}=0.
\ee There exists one class of solutions to these equations for $s<
0.769$:  the Maclaurin spheroids with
$a=b$ and an eccentricity $e=(1-{c^2\oo a^2})^\half<0.813$. For
$s\rightarrow 0$, ${a\oo c}={b\oo c}\rightarrow 1$; spheroids turn
into spheres of infinite radius and zero mass.
 When
$s> 0.769$  there are two classes  of solutions: Maclaurin spheroids with
$e>0.813$ and Jacobi ellipsoids with non equal axis. At $s= 0.769$
there is thus a bifurcation.

The solutions are shown and discussed in details in Chandrasekhar's
book together with  their stability.  Here we use the topology of
linear series with conjugate variables with respect to minus the
potential energy to find stability conditions.
\vskip  .5 cm (b) Stability conditions. 
\vskip  .5 cm Let $F_e(s)$ represent the extremal values of
$F$. $F_e(s)$ is equal to minus the potential energy times a constant.
Equilibrium configurations are thus stable if $F_e(s)$ is a maximum of
$F$. The conjugate parameter of
$s$ with respect to $F_e(s)$ is
\be K(s)\equiv{dF_e(s)\oo ds}=-{(X^1X^2)^{1\oo 3}\oo X^1+X^2}.
\ee The linear series reproduced  from \cite{K79} is shown in figure
4. We see the bifurcation appearing at point
$B$ where  $s= 0.769$. Since homogeneous spheres are stable
\cite{LY} Maclaurin spheroids are stable for $0<s<0.769$ and
$0<e<0.813$. A change of stability can only occur at point $B$ as we
know but we cannot tell what happens at a bifurcation point just by
looking at the linear series.

Let us then look as suggested in (\ref{+epsilon}) at solutions of
modified equations like, for instance,
\be {\di V\oo\di x^1}=\epsilon ~~~{\rm and } ~~~{\di V\oo\di x^2}=0,
\ee with $\epsilon$ small. The effect of $\epsilon$ is to break the
symmetry and lift the bifurcation. Figure 4 represents also the linear
series $K(s) $ for the perturbed solutions with
$\epsilon=-0.01$. Any other small value of $\epsilon$ would have a
similar effect on the topology but curves  with 
$\epsilon>0$ cut those with $\epsilon=0$ and the drawing is not so
nice.  We now see a continuous line (sequence 1) which connects
smoothly for
$\epsilon\rightarrow 0$ to the linear series of stable Maclaurin
configurations for
$s<0.769$ and to Jacobi ellipsoids which exists only for $s>0.769$.
Sequence 1 has no vertical tangent. Thus by continuity  Jacobi
ellipsoids must be stable. Sequence 2 has a vertical tangent at point
$C$. The $CD$ branch   identifies  with the stable sequence 1 when
$\epsilon\rightarrow 0$,  and the $CE$ branch represents necessarily
unstable configurations with one unstable mode\footnote{The two
Poincar\'e coefficients of stability are different at point $B$, see
\cite{K79}.}. Taking $\epsilon\rightarrow 0$, we conclude that
Maclaurin spheroids with eccentricities 
$e>0.813$  are unstable.

The stability limits have been known since Poincar\'e's time
\cite{P02}. The  instability of Maclaurin spheroids for
$e>0.813$ is  secular; it
   shows up with a small viscous dissipation in a dynamical
perturbation calculation \cite {Chandra69}.
\vskip  .5 cm
\normalsize


\nnn {\it (iii) Final comments and more examples.}
\vskip  .5 cm
 (a) Final Comments
\vskip  .5 cm 
Thermodynamics of self-gravitational systems is helpful
for calculating statistical equilibrium configurations, thermodynamic
functions, stability limits and the effect of fluctuations near
instability. 

Globular clusters and other astronomical objects are only close to
statistical equilibrium and  evolve more often slowly. When evolution
is slow enough these objects pass through a series of quasi-equilibrium
configurations with ever increasing entropy. In this respect, linear
series also mimic evolutionary tracks. Turning points indicate the
limits beyond which evolution proceeds in a non isothermal way,
generally towards higher central densities and core collapse. The
fluctuations move somewhat the stability limits to lower temperatures. 

What happens next, once thermal equilibrium no more exists, is of
immense interest but does not belong to this paper. Thermodynamics is 
a small chapter in theoretical astronomy  though an    interesting one.
\vskip  .5 cm 
(b) More applications
\vskip  .5 cm 
Thermodynamic methods as described in the present work
have been used in various studies which we shall now mention briefly.
One dimensional isothermal parallel sheets have been shown to be
stable to one dimensional perturbations \cite{KL}. Isothermal
axially-symmetric equilibria of gravitating rods are generally stable
with respect to cylindrically symmetric perturbations \cite{KL78}.
There is no gravothermal catastrophe as in isothermal spheres but no
equilibria exist below some finite temperature. In contact with a heat
bath slightly colder than that, the system collapses slowly, giving up
an unlimited amount of energy. The effect of a mass spectrum on the
stability limits of isothermal spheres has also been analyzed
\cite{KT}.

 Isothermal spheres with a Fermi-Dirac distribution behave in some
ways like  systems of  hard  spheres particles: they have  phase
transitions from gaseous to core-halo structure.  Stability limits
with  a Fermi-Dirac distribution were analyzed by Chavanis and Sommeria
\cite{Chav98}, see also \cite{Chav}. A cutoff in potential has a
similar effect \cite{AH}. Another example in which there may be phase
transitions   from  white dwarfs to neutron stars but where the
turning point method has {\it not} been used, though it would have
proven extremely useful, is the study by Harrison, Thorne Wakano and
Wheeler \cite{HTWW} of cold catalyzed matter. They minimize the
mass-energy $M$. The only independent parameter is the number of
baryons $N$. A plot of the conjugate parameter $\di (-M)/\di N$ versus
$N$ using their own table on page 152-153 would have shown   the
stability conditions and  in particular the number of unstable modes
in one stroke
\cite {K80/2} as is clear from figure 5.

Chavanis \cite{Chav2} also studied isothermal spheres in Newtonian
mechanics but with a relativistic equation of state. He found that
instability sets in at smaller density contrasts but stronger binding
energies.

Stability conditions in  relativistic  spheres   with  energy cutoffs
\cite {KHK}\cite{HK2} have also {\it not} been treated  by the 
turning point method but it is obvious from  figures in \cite{KHK}
that the their results
 might have been obtained in a more simple way.

Since Bekenstein \cite{BE}\cite{BE2} attributed an entropy to black
holes and Hawking
\cite {HA74} found that they emit  black body radiation of quantum
origin, studies of their stability in various surroundings and with
various charges have flourishes. After the seminal papers by Hawking  
\cite{HA76} and by Gibbons and Perry\cite{GP} on the thermodynamics
of  black holes, there appeared a series of papers based on linear
series devoted to Kerr black holes
\cite{KOK} and Kerr-Newman black holes \cite{KOK2}. A more detailed
analysis of black holes in a cavity which takes account of the effect
of fluctuations started in
\cite{OKP} was  analyzed in greater details in \cite{PKO}. All those
papers neglect the mass-energy of the cavity, usually a thin shell,
that is supposed to play the role of a massless  reflective wall.
It has been pointed out   \cite{OKK} that thin shells which do not
allow infinite tensions\footnote{Thin shells that satisfy the dominant
energy condition
$-\rho<P<\rho$ with $\rho>0$.} must contain at least $30\%$ of the
total  mass-energy. This may affect considerably the stability limits
and throws some doubts on considerations, with sometimes far reaching
consequences, that were derived from black holes in equilibrium with
radiation in weightless cavities
\cite{SWZ}.


\vskip .5 cm
{\Large {\bf {Acknowledgements}}}
\vskip .5 cm
This work grew out of a series of lectures at the Laboratoire de Physique
Th\'eorique at the University of Orsay in November 2001. The lectures
entitled ``Thermodynamique des gaz autogravitants" covered a great deal
more material than the present much centered work. I wish to thank
Jean-Philippe Uzan who gave me the opportunity to rethink  this
interesting subject. Useful conversations ensued with many people. I 
thank in particular Nathalie Deruelle and Bernard Jancovici whose
clarifying remarks were particularly helpful. 

Donald Lynden-Bell read a first draft of this paper and made highly
valuable comments which were incorporated in the text. Pierre-Henri
Chavanis who read the second draft contributed additional useful
remarks which I also incorporated. 

I am also grateful to Evgeni Sorkin, Ehud Nakar and Jonathan Oren for
their help with the cosmetics of the figures.

I thank the Laboratoire de Physique Th\'eorique at Orsay for a warm
welcome.


\vskip .5 cm
\centerline{\Large{\bf{Appendices}}}
\begin{appendix}
\setcounter{equation}{0}
\renewcommand{\theequation}{\Alph{section}.\arabic{equation}}
\section{Proof that (\ref{Om3}) can be
transformed into (\ref{Om4})}
\normalsize

We start from (\ref{Om3})
with $H=H_K+H_P$ defined in (\ref{H1}). The trick consists in
introducing first a ``continuous density" $\rho ({\bf r})$:
\be
\rho({\bf r})=\sum\limits_{i}m\delta({\bf r }- {\bf r}_i).
\label{rho1} 
\ee The   potential energy $H_P$ can then be written
\be H_P=-{1\over 2}\int_V{G\rho({\bf r})
\rho({\bf r'}) \over |{\bf r}-{\bf r'}|}d^3rd^3r'.
\label{HP1}
\ee Next we discretize  the whole space, cutting it  into small cubes
with an indice $a$ and volume
$\kappa ^3$. Replacing 
$\rho ({\bf r})d^3r$ in (\ref{HP1}) in each  cube with indices 
$a$ by $\rho^a\kappa ^3$ the potential energy takes
 the form
\be H_P={1\over 2}\sum\limits_{a,b} u_{ab}\rho^a\rho^b.
\label{HP2}
\ee   The matrix $u_{ab}$ is the discrete inverse of the Laplacian and
more precisely
\be
\lim_{a,b\rightarrow\,{\rm continuum}} u^{-1}_{ab}={\Delta \over 4\pi
G}\delta({\bf r}-{\bf r}').
\label{u-1}
\ee Let $W_a$ represent the mean field in discretized form. Define a
set of variables $X_a$ to be integrated out soon:
\be X_a= W_a-\beta^{1/2}u_{ab}\rho^b.
\label{XA}
\ee  
The following identity\footnote{This old identity due to
Stratonovich \cite{St} has proved useful in many-body physics.}  is readily
constructed with (\ref{HP2}) and  (\ref{XA}) by elementary algebraic
manipulations:
\be -\beta H_P+{1\over 2}u^{-1ab}X_aX_b=-\beta^{1/2}W_a\rho^a+{1\over
2}u^{-1ab}W_aW_b.
\ee Then taking the exponent of both sides of this expression and
integrating over the whole domain of variation of the
$W_a$'s and
$X_a$'s leads to:
\be e^{-\beta H_P}={1\over [{\rm det}(-2\pi u_{ab})]^{{1\over
2}}}\int_{-\infty}^{+\infty}e^{\lll-\beta^{1\over 2}W_a\rho^a+{1\over
2}u^{-1ab}W_aW_b\rrr}\prod\limits_a dW_a.
\ee We now return  to continuity, taking into account   formula
(\ref{HP1}) and (\ref{u-1}); this   lets us write
$e^{-\beta H_P}$ in the following form:
\be e^{-\beta H_P}={1\over\bb}\int_{-\infty}^{+\infty}e^{\left
[-m\sum\limits_i\beta^{1\over 2}W({\bf r}_i) +{1\over 8\pi
G}\int_{-\infty}^{+\infty}W( {\bf r})\Delta W({\bf r})d^3r\right
]}{\cal D}W,
\label{e-HP1}
\ee
 where
\be {\cal B}= \lim_{a,b\to\rm continuum}\sqrt{det(-2\pi u_{ab})}
\quad\hbox{and}\quad {\cal D}W=\lim_{a\to\rm continuum}\prod_a dW_a.
\ee
$\bb$ is certainly divergent; $u_{ab}$ tends to $1/|{\bf r}-{\bf
r}'|$. We assume that a short distance cutoff will make $\bb$
convergent. The nature of the cutoff is unimportant in our
approximation - see section 1(iii)(c). The  expression for
$e^{-\beta H_P}$  obtained in (\ref{e-HP1}) we put back  into
$\Om(E)$ in (\ref{Om3}). It  takes now little work to see that
$\Om$ can   be written as follows
\be
\Om = {1\over (2\pi i)\bb}\int_{b-i\infty}^{b+i\infty}
\int_{-\infty}^{+\infty}e^{\LLL\beta E+{1\over 8\pi
G}\int_{-\infty}^{+\infty} W({\bf r})\Delta W({\bf r})d^3r\RRR}\Psi
d\beta{\cal D}W,
\label{Om5}
\ee  where
\be
\Psi ={1\over N!}\left(\int e^{-\beta [{1\over 2m} p^2
+m\beta^{-{1\over 2}}W({\bf r})]}d\om\right)^N~~~{\rm is~ of~
the~form}~~~\Psi ={1\over N!}\psi^N.
\ee  
This is the of the same form as $\td\Psi$ in (\ref{Psi11}). We may thus
replace $\Psi$ with the same approximate expression as the one
obtained for $\tilde
\Psi$ in (\ref{Psi11}) to (\ref{Psi12}) :
\be
\Psi={1\over N!}\psi^N\simeq{1\over 2\pi
i}\int_{a-i\infty}^{a+i\infty}\LLL\sum\limits_{N'=0}^\infty {1\over
N'!}\psi^N{'} e^{\al (N'-N)}\RRR d\al={1\over 2\pi
i}\int_{a-i\infty}^{a+i\infty}
\LLL\sum\limits_{N'=0}^\infty {1\over N'!}f^N{'}\RRR e^{-\al N}d\al,
\label{Psi}
\ee  in which
\be f=e^{\al-\bt \LLL{1\over 2m} p^2+m\bt^{-{1\over 2}}W({\bf r})\RRR}.
\ee
 We can also write $\Psi$ like this
\be
\Psi={1\over 2\pi i}\int_{a-i\infty}^{a+i\infty} e^{\lll-\al N+\int
fd\om\rrr}d\al;
\ee replacing $\Psi$ by this expression in (\ref{Om5}) gives 
$\Om(E)$ as written in equation (\ref{Om4}) with $\si$ shown in equation
(\ref{si}).

 We followed closely the calculations\footnote{Paper
\cite{HK1} contains correct calculations but  the wrong
interpretation. The authors believed they were dealing with isolated
systems when they dealt in fact with a grand canonical ensemble.} in
\cite{HK1} and \cite{HK} . We left aside all sorts of intricacies
about   existence and convergence problems in functional integrations.
These were given more  carefully considerations in \cite{HK1} with
appropriate references.


\setcounter{equation}{0}
\section{Calculation of $\Om(E)$ to order two
in $\delta W $ - Formula (\ref{Om19})}
We start from
$\si(\al,\bt,W)$ given in equation (\ref{si}) which we expand in a
Taylor series to order two in $\de\al=\al-\al_e, \de \bt=\bt-\bt_e, 
 \de W=W-W_e$ around their extremal values $\al_e, \bt_e, W_e$
defined by (\ref{Poisson})-(\ref{NE}). There is no reason \`a priori to limit
$\delta W$ to continuous perturbations except for the fact
\cite{GI} that non-continuous functions form a subset of measure zero.
Here are the first order derivatives:
\ba {\di\si\over\di \al}&=&-N+\int f d\om   ~~~,~~~{\di\si\over\di
\bt}= E-\int
\left[{1\over 2m}p^2+{1\over 2}m\bt^{-\half}W({\bf r})\right]f
d\om,\nonumber\\
\int_{-\infty}^{+\infty}{\di\si\oo\di W}\delta Wd^3r&=&
-m\bt^\half\int\delta Wfd\om+{1\oo 4\pi
G}\int_{-\infty}^{+\infty}\delta W\Delta Wd^3r.
\ea  It will be of great help to define with an over-bar the mean
values in the phase space of one particle. For instance:
\be
\Bar{X}={1\oo N}\int Xf_e d\om
\ee  Thus, the mean value of the total energy, equation (\ref{NE}), can
be written as follows:
\be  E=E_K+E_P=N\Bar{\lll {p^2\oo 2m}\rrr}+\half mN\Bar{U}=N\lll {3\oo
2\bt_e}+{\half m\Bar{U}}\rrr~~\Rightarrow~~ {\bt_e E\oo N}={3\oo
2}+\half m\bt_e \Bar{U}.
\label{Energy}
\ee This expression will soon be useful. Another useful relation is
the virial equality:
\be  E+E_K=3P_bV ~~~\stackrel{(\ref{P})}\Longrightarrow~~~{\bt_e E\oo
N}+{3\oo 2}=3{\rho_b\oo (mN/V)}=3{\rho_b\oo \rho_{mean}}.
\label{Virial}
\ee

Here now are the second derivatives of $\si$ with respect to
$\al,\bt,W$ calculated at the point of extremum:
\ba
\lll {\di^2\si\oo \di^2\al}\rrr_e =N~~~&,&~~~\lll {\di^2\si\oo
\di\al\di\bt}\rrr_e
 = -E~~~,\nonumber\\
\lll {\di^2\si\oo \di^2\bt}\rrr_e &=& {N\oo
4}\LLL{15\oo\bt_e^{2}}+{7m\oo\bt_e}\Bar{U}+m^2(\Bar{U})^2+m^2\Bar{\lll
U-\Bar{U}\rrr^2} \RRR,\nonumber\\
\int\lll {\di^2\si\oo \di\al\di W}\rrr_e\delta Wd^3r&=&
-Nm\bt_e^\half\Bar{\delta W},\nonumber\\
\int\lll {\di^2\si\oo \di\bt\di W}\rrr_e\delta Wd^3r&=&N\LLL
m\bt_e^{-\half}\Bar{\delta W}+\half m^2\bt_e^\half\Bar{(U\delta
W)}\RRR,\nonumber\\
\int\int\delta W\lll {\di^2\si\oo \di W\di W'}\rrr_e\delta
W'd^3rd^3r'&=&Nm^2\bt_e\Bar {(\delta W)^2}+{1\oo 4\pi G}\int\delta
W\Delta\delta Wd^3r.
\ea 
With these derivatives we may write the expression for
$\delta^2\si$ at the extremum:
\ba  
2\delta^2\si & = & \underbrace{N(\delta\al)^2}_1 +
\underbrace{(-2E\delta\al\delta\bt)}_2 \nonumber \\  &+ &
\underbrace{(N/4)\LLL 
15\bt_e^{-2}+7m\bt_e^{-1}\Bar{U}+m^2(\Bar{U})^2+m^2\Bar{(U-\Bar{U})^2}
\RRR(\delta\bt)^2}_3\nonumber
\\ &+& \underbrace{(-2Nm\bt_e^\half\delta\al\Bar{\delta
W})}_4+\underbrace{N\LLL  2m\bt_e^{-\half}\Bar{\delta
W}+m^2\bt_e^\half\Bar{(U\delta
W)}\RRR\delta\bt}_5+\underbrace{Nm^2\bt_e\Bar{(\delta
W)^2}}_6\nonumber \\ &+&
\underbrace{(4\pi G)^{-1}\int_{-\infty}^{+\infty}\delta W\Delta\delta
Wd^3r}_7.
\ea  
We now use (\ref{Energy}) and (\ref{Virial}) to simplify some of
the terms. {\it Our aim is to reduce $\delta^2\si$ to a sum of squares
that can be separately integrated in $\al,\bt$ and
$W$ or equivalently in $\delta\al,\delta\bt$ and  $\delta W$}. We
assume as usual that the exponent is steep and that order two is a
good approximation.
$\delta\al$ and
$\delta\bt$ vary along a line parallel to the imaginary axis between
$\pm\infty$.

First we write
\be
\underbrace{{\rm Terms}}_{1+2+3}= \underbrace{N\lll \delta\al - {\bt
E\oo N}{\delta\bt\oo\bt_e}\rrr^2 }_{1'}+\underbrace{ Nb^2\lll \delta
\bt\oo \bt_e\rrr^2 }_{2'}~\Rightarrow~b^2={1\oo 4}\LLL  {6\rho_b\oo
\rho_{mean}}+m^2\bt_e^2\Bar{(U-\Bar{U})^2}\RRR>0.
\ee Second, 
\ba
\underbrace{{\rm Terms}}_{4+5}&=&\underbrace{-2Nm\bt_e^\half\lll
\delta\al - {\bt_e E\oo N}{\delta\bt\oo\bt_e}\rrr\Bar{\delta W}
}_{3'}+\underbrace{2N\Bar{(g\delta W)}{\delta\bt\oo
\bt_e}}_{4'},\nonumber\\ {\rm where}~~~\Bar{(g\delta
W)}&=&m\bt_e^\half\lll  1-{\bt_e E\oo N}\rrr\Bar{\delta W}+ \half
m^2\bt_e^{3\oo 2}\Bar{(U\delta W)}.
\ea  With these two results we can now see that the quadratic sum
$\delta^2\si$ reduces to this:
\ba  2\delta^2\si &=& \underbrace{N\lll \delta\al - {\bt_e E\oo
N}{\delta\bt\oo\bt_e}\rrr^2 }_{1'}+\underbrace{ Nb^2\lll \delta \bt\oo
\bt_e\rrr^2 }_{2'}+\underbrace{\LLL-2Nm\bt_e^\half\lll \delta\al -
{\bt_e E\oo N}{\delta\bt\oo\bt_e}\rrr\Bar{\delta W}
\RRR}_{3'}\nonumber\\ &+&\underbrace{2N\Bar{(g\delta
W)}{\delta\bt\oo\bt_e}}_{4'}+\underbrace{Nm^2\bt_e\Bar{(\delta
W)^2}}_6+\underbrace{(4\pi G)^{-1}\int\delta W\Delta\delta Wd^3r}_7.
\label{113}
\ea  The sum of  $1'$ and $3'$, it is a difference of positive
quantities:
\ba
\underbrace{{\rm Terms}}_{1'+3'}&=&N\lll \delta\al - {\bt_e E\oo
N}{\delta\bt\oo\bt_e}\rrr^2 -2Nm\bt^\half\lll \delta\al - {\bt_e E\oo
N}{\delta\bt\oo\bt_e}\rrr\Bar{\delta W} \nonumber\\ &=& N\lll 
\delta\al - {\bt_e E\oo N}{\delta\bt\oo\bt_e}-m\bt_e^\half\Bar{\delta
W}\rrr^2-Nm^2\bt_e(\Bar{\delta W})^2,
\label{1'3'}
\ea  
next we sum  $2'$ and $4'$ which is of the same type as $1'+3'$:
\be
\underbrace{{\rm Terms}}_{2'+4'}=Nb^2\lll {\delta
\bt\oo\bt_e}\rrr^2+2N\Bar{(g\delta W)}{\delta \bt\oo\bt_e}=N\LLL 
b{\delta
\bt\oo\bt_e}+{\Bar{(g\delta W)}\oo b}\RRR^2-N\LLL{\Bar{(g\delta W)}\oo
b}\RRR^2.
\label{2'4'}
\ee  We now observe that $2\delta^2\si={\rm eq.} ( \ref{1'3'})+{\rm
eq.} (\ref{2'4'})+{\rm Terms} (6+7)$. Let us introduce a couple of {\it
real} variables which vary between
$\pm\infty$, similar to those introduced in equations (\ref{idelta}):
 \be 
 i\al^*=\delta\al - {\bt_e\oo
N}{\delta\bt\oo\bt_e}-m\bt_e^\half\Bar{\delta W}~~~{\rm
and}~~~i\bt^*= b{\delta
\bt\oo\bt_e}+{\Bar{(g\delta W)}\oo b}.
\ee
 In terms of $\al^*$, $\bt^*$ and $\delta W$, $\delta^2\si$ reduces to 
\ba
\delta^2\si&=& -\half N(\al^{*2}+\bt^{*2})+ \delta ^2\si' ~~~{\rm
in~which}\nonumber\\
\delta ^2\si'&=&{1\oo 8\pi G}\int\delta W\Delta \delta W d^3r+ {N\oo
2}\LLL  m^2\bt_e\Bar{(\delta W-\Bar{\delta
W})^2}-b^{-2}\{{\Bar{(g\delta W)}\}^2}\RRR.
\ea  With this result we can calculate $w$ to order two starting from
(\ref{Om4}), (\ref{si}); inserting $\si=S+\delta^2\si$ into
(\ref{Om4}) we obtain
\be 
\Om(E)\simeq {1\oo (2\pi
i)^2{\bb}}\int_{-\infty}^{+\infty}\dd
W\int_{-\infty}^{+\infty}\int_{-\infty}^{+\infty}e^{S+\delta^2\si}d\al
d\bt={\bt_e e^S\oo 2\pi bN}\int_{-\infty}^{+\infty}\dd
We^{\delta^2\si'}.
\ee  
 This is equivalent to formula (\ref{Om19}). 

It is worth noticing that $\delta^2\si$, see (\ref{113}), is equal to
 terms $(6+7)$ minus two squares, see (\ref{1'3'}) and (\ref{2'4'}).
This implies that:
\be {\delta^2\si'}\leq {\rm Terms}~(6+7)= \half
\int_{-\infty}^{+\infty}\delta W\lll{\Delta\oo 4\pi G}+m\bt_e
\rho\rrr\delta W d^3r={\delta^2\si}_{GC}
\ee
${\delta^2\si}_{GC}$ is the second order term of the exponent for the
grand canonical ensemble $\si_{GC}$ introduced in  equation
(\ref{GCE}). The spectrum of eigenvalues of this quadratic form has
been studied in detail in \cite{HK1} for isothermal spheres  where it
was shown  that in a stable system 
${\delta^2\si}_{GC}\rightarrow-\infty$. Thus
${\delta^2\si}\rightarrow-\infty$ as well for such configurations.
See section 1(iii)(c) about this divergence. 
\end{appendix}

\vskip .5 cm
\centerline{\nnn\Large{\bf{Figure captions}}}
\normalsize
\vskip .5 cm

{\bf Figure 1}: This figure illustrates the point made in section 4
subsection {\it (i)} where full details are given.

{\bf Figure 2}:  The figure represents the pair of conjugate
thermodynamic functions $\bt(E)$ in appropriate units for isothermal
spheres described in section 4{\it (i)}$(\al)$. The linear series gives
stability limits in either isolated spheres or spheres in a heat bath. The
parameter along the line, at vertical and at  horizontal tangents is the
density contrast $\rr=\rho(center)/\rho(boundary)$.

{\bf Figure 3a and 3b}: These figures represent the same pair of conjugate
thermodynamic functions $\bt(E)$ as in figure 2 with different units
(defined in the text) and is appropriate to detect stability limits  of
isothermal spheres in a grand canonical or a grand
micro-canonical ensemble as described in section 4{\it (i)}$(\bt)$. Figure
3a is for $-0.008<E<0$. Figure 3b is for
$0<E<0.6$. For density contrasts $\rr\rightarrow 1$, $E\rightarrow 0$
and
$\bt\rightarrow \infty$. Thus the line with density contrasts $\rr>1.66$
will, before $E$ turns negative, cuts the line that comes from infinity . 

{\bf Figure 4}: This figure represents the pair of conjugate variables for
Maclaurin and Jacobi ellipsoids described in 4{\it (ii)}. The figure is
reproduced from \cite{K79} in which our small $s$ was represented by
$S$. $S$  here  represents  the entropy.

{\bf Figure 5}: This is a topologically correct plot of 
$\di(-E)/\di N$ as a function of the number of baryons in units of
$N_\odot=10^{57} $. The parameter along the line is the central density
in $gr\cdot cm^{-3}$. The linear series shows six consecutive turning
points calculated by Harrison. There are two stable branches
corresponding respectively to cold white dwarfs and to neutron stars.
More details are given in \cite{HTWW}.  

\end{document}